\documentclass[twocolumn]{aastex631}

\usepackage{xcolor}

\newcommand{\hi}{{\rm H\,{\small I}}}

\newcommand{\kms}{\ensuremath{{\rm km\,s^{-1}}}}
\newcommand{\percc}{\ensuremath{{\rm cm^{-3}}}}
\newcommand{\persc}{\ensuremath{{\rm cm^{-2}}}}
\newcommand{\fmol}{\ensuremath{f_{\mathrm{mol}}}}
\newcommand{\hcop}{\text{HCO\textsuperscript{+}}}

\newcommand{\htwo}{\text{H\textsubscript{2}}}

\newcommand{\texc}{\ensuremath{T_{\rm{ex}}}}
\newcommand{\ts}{\ensuremath{T_{\rm{s}}}}
\newcommand{\tcmb}{\ensuremath{T_{\rm{CMB}}}}
\newcommand{\xco}{\ensuremath{X_{\rm{CO}}}}

\newcommand{\codarkb}{CO-dark*}
\newcommand{\codarka}{CO-dark}

\begin{document}

\title{CO-dark molecular gas traced by \hcop{} in the diffuse interstellar medium}

\correspondingauthor{Daniel R. Rybarczyk}
\email{rybarczyk@astro.wisc.edu}

\author[0000-0003-3351-6831]{Daniel R. Rybarczyk}
\affiliation{University of Wisconsin--Madison, Department of Astronomy, 475 N Charter St, Madison, WI 53703, USA}

\author[0000-0003-4961-6511]{Michael P. Busch}
\altaffiliation{Jansky Fellow of the National Radio Astronomy Observatory}
\affiliation{National Radio Astronomy Observatory, 520 Edgemont Road, Charlottesville, VA 22903, USA}

\author[0000-0003-0235-3347]{J. R. Dawson}
\affiliation{School of Mathematical and Physical Sciences and Astrophysics and Space Technologies Research Centre, Macquarie University, Sydney 2109, Australia}
\affiliation{Australia Telescope National Facility, CSIRO Space \& Astronomy, PO Box 76, Epping, NSW 1710, Australia}

\author[0000-0002-9888-0784]{Min-Young Lee}
\affiliation{Korea Astronomy and Space Science Institute, 776 Daedeok-daero, Daejeon 34055, Republic of Korea}
\affiliation{Department of Astronomy and Space Science, University of Science and Technology, 217 Gajeong-ro, Daejeon 34113, Republic of Korea}

\author[0000-0002-1583-8514]{Gan Luo}
\affiliation{Institut de Radioastronomie Millimetrique, 300 rue de la Piscine, 38400, Saint-Martin d’Hères, France}

\begin{abstract}
A classic problem in the study of the interstellar medium (ISM) is the near-invisibility of molecular hydrogen (\htwo{}) in cold environments. Observations of CO emission are typically used to indirectly trace \htwo{}, but a significant fraction of \htwo{} in the diffuse ISM is not associated with any detectable CO emission (``CO-dark'' molecular gas). Meanwhile, observations of \hcop{} absorption trace nearly all of the \htwo{} in diffuse directions. In particular, a kinematically broad \hcop{} absorption signature traces extremely diffuse, CO-dark \htwo{}. We have used sensitive observations of \hcop{}, CO, and atomic hydrogen (\hi{}) in absorption to constrain the properties of such diffuse molecular gas in five directions. The diffuse molecular gas revealed by broad \hcop{} absorption has a lower fraction of cold \hi{} ($f_{\mathrm{CNM}} = 0.38^{+0.28}_{-0.27}$) and a lower fraction of hydrogen in \htwo{} ($f_{\mathrm{mol}}=0.09^{+0.06}_{-0.03}$) than gas traced by CO in the same directions. 
We detect almost no CO absorption from the gas traced by broad \hcop{} absorption. We constrain the CO abundance relative to \htwo{} to be $\lesssim10^{-6}$--$10^{-5}$ for gas traced by both broad and narrow \hcop{} absorption, consistent with chemical model predictions for the diffuse ISM. We further show that neither CO emission nor absorption is likely to be detected where $N(\htwo{})\lesssim\mathrm{few}\times10^{19}~\persc{}$ --- a result of both the low CO abundance and the low H$_2$ column --- while \hcop{} absorption is readily detected for $N(\htwo{})\gtrsim\text{few}\times10^{18}~\persc{}$. 
These results demonstrate that even modest amounts of cold \hi{} can bear \htwo{}, providing critical constraints on the \hi{}-to-\htwo{} transition in the ISM.

\end{abstract}

\section{Introduction} \label{sec:intro}
Molecular hydrogen (\htwo{}) is the most abundant molecule in the interstellar medium (ISM) and plays a key role in chemistry and star formation \citep[e.g.,][]{Bigiel2008}. Yet, because \htwo{} lacks a permanent dipole moment and because of the large energy spacing between the two lowest rotational transitions, it is extremely difficult to observe directly at low temperatures. Instead, alternative tracers in emission and absorption are used to investigate the distribution, kinematics, and structure of the molecular ISM.

The CO molecule is one of the most commonly used tracers of the molecular content of galaxies because of its high abundance\footnote{Throughout the text, we refer to the ``abundance'' of different molecular species as the ratio between molecular column density to the \htwo\ column density, unless otherwise specified.} and its brightness in molecular clouds. Yet, the relationship between the integrated CO emission and the \htwo{} column density, summarized with the \xco{} factor, $\xco{}=N(\htwo{})/W_{\mathrm{CO}}$, varies between and within galaxies \citep[see review by][and references therein]{Bolatto2013}. 

Additionally, a significant quantity of ``CO-dark'' molecular gas (that is, \htwo{} not traced by CO emission) exists in diffuse ($A_V\lesssim1~\mathrm{mag}$) environments \citep[e.g.,][]{Allen2015,Liszt2019,Busch2019,Busch2021,LisztGerin2023b,Luo2024}. 
Two important tracers of this diffuse molecular component are OH emission and \hcop{} absorption
\citep{Liszt2019,Busch2021,LisztGerin2023b}. Since both OH and \hcop{} are known to have nearly fixed abundances with respect to \htwo{} in diffuse environments \citep{LisztLucas2000,XuLi2016,Nguyen2018,LisztGerin2023a}, they are considered to be excellent tracers of diffuse molecular gas.

Intriguingly, a kinematically-broad CO-dark spectral signature has recently been detected in both OH emission \citep{Busch2021} and \hcop{} absorption \citep{LisztLucas2000,Rybarczyk2022_Obs}. The molecular gas traced by these signatures has different temperature and density properties than ``CO-bright'' gas detected along the same lines of sight \citep{LisztLucas2000,LisztPety2012,Busch2021,Rybarczyk2022_Obs}. 
\citet{Busch2021} showed that broad OH emission traces large-scale Galactic velocity structure. Specifically, they argued that the OH emission was evidence for a thick ($\sim200~\mathrm{pc}$) disk of CO-dark diffuse molecular gas in the Outer Galaxy, invisible to canonical Galactic CO surveys like \citet{DameThaddeus2001}. 
Meanwhile, \citet{Rybarczyk2022_Obs} argued that the molecular gas traced by broad \hcop{} absorption probes the earliest stages of molecule formation in the Galactic ISM, where the molecular fraction of hydrogen, $\fmol=2n_{\htwo}/n_{\mathrm{H}}$, is $\lesssim10\%$.
Interestingly, previous independent analyses of these two signatures have suggested that both could be explained 
by the presence of molecular gas with a mean \htwo{} number density of $n\sim10^{-3}$--$10^{-2}~\percc$ \citep{LisztLucas2000,Busch2021}, although observations do not rule out the presence of unresolved higher-density clumps.
These observations provide valuable constraints on the physical properties of the reservoir of CO-dark molecular gas that exists in the diffuse ISM.

Atomic hydrogen (\hi{}) provides the raw material for the formation of \htwo{} \citep[which occurs primarily on the surfaces of dust grains; see][and references therein]{Cazaux2002}, so characterizing the properties of \hi{} is key to understanding how molecule formation commences. 
\hi{} exists as a multiphase medium, with gas distributed between a warm, diffuse phase (the warm neutral medium, ``WNM''); a colder, denser phase (the cold neutral medium, ``CNM''); and a thermally unstable phase at intermediate temperature and density \citep[the unstable neutral medium, ``UNM;'' e.g.,][and references therein]{21SPONGE_2018,McG2023}. The CNM is particularly important in fueling molecule formation, since it provides the number and column densities necessary to (a) form molecules at an appreciable rate, and (b) shield them from photodissociation. It is unsurprising, therefore, that molecular gas is found preferentially in directions with a high fraction of \hi{} in the form of CNM \citep[][]{SS2014,Nguyen2019,Rybarczyk2022_Obs,Park2023,Hafner2023}.

Yet, \citet{Rybarczyk2022_Obs} showed that the \hi{} associated with broad, CO-dark \hcop{} absorption has a systematically lower CNM fraction than the \hi{} associated with CO-bright molecular gas, 
suggesting that molecules can form in environments with significantly lower CNM column densities than previously thought \citep[e.g.,][]{SS2014}. 
In fact, in some directions the broad CO-dark \hcop{} signature traces nearly all velocities where \hi{} is detected in absorption --- a result that was also found in the earlier work of \citet{LisztLucas2000}. This is in contrast to the behavior of CO-bright gas, which is generally coincident only with the coldest, optically thickest \hi{} \citep{Nguyen2019,Park2023,Hafner2023}. 
Even more surprising, \citet{Busch2021} showed that, in certain directions, broad OH emission traces nearly all velocities where \hi{} is detected in \textit{emission}. Whereas \hi{} absorption arises primarily from the CNM (and some UNM), the \hi{} emission arises from a mixture of CNM, UNM, and WNM.

Despite the intriguing results presented by \citet{LisztLucas2000}, \citet{Busch2021}, and \citet{Rybarczyk2022_Obs}, these investigations of the CO-dark diffuse molecular gas were limited in scope. For example, because \citet{Busch2021} investigated the atomic gas in the diffuse ISM using only \hi{} emission, they were unable to investigate the multiphase \hi{} properties in detail --- the CNM, UNM, and WNM can only be reliably separated using the combination of \hi{} emission and absorption \citep[e.g.,][]{HT03}.
Meanwhile, the classification of CO-dark gas used by \citet{Rybarczyk2022_Obs} was based on CO emission data from \citet{DameThaddeus2001}, but the definition of ``CO-dark'' always implicitly depends on CO sensitivity \citep[e.g.,][]{Donate2017}. Indeed, \citet{Li2018} showed that more sensitive CO emission observations often detected diffuse CO where \citet{DameThaddeus2001} did not, leaving open the possibility that the non-detections in \citet{Rybarczyk2022_Obs} were merely a consequence of  sensitivity. 
Moreover, the \citet{Rybarczyk2022_Obs} \hcop{} absorption observations probed pencil-beam sightlines, which trace much smaller physical scales on the plane of the sky than the relatively large CO emission beams, so their CO and \hcop{} observations were not necessarily probing the same molecular gas structures. 
Finally, \citet{LisztLucas2000} did not compare their broad \hcop{} absorption observations to observations of CO in the same directions. 
Therefore, high-sensitivity observations of \hcop{}, CO, and \hi{} (in absorption) at a comparable spatial resolution are essential to quantify the physical properties and evolution of diffuse molecular gas.

\begin{deluxetable*}{c|c|c|c|c}[] \label{tab:categories}
\tablecaption{Categories of molecular line observations}
\tablehead{
\colhead{Category} & \colhead{\hcop{} absorption} &  \colhead{CO absorption} & \colhead{CO emission} & \colhead{Description}
}
\startdata
    1 & No & No & No & Atomic \\
    2 & Yes & No & No & \codarka \\
    3 & Yes & Yes & No & \codarkb \\
    4 & Yes & Yes & Yes & CO-bright \\
\enddata
\end{deluxetable*}

In this work, we directly compare sensitive observations of \hcop{}, CO, and \hi{} in absorption in five directions where broad \hcop{} absorption has previously been detected \citep{Luo2020,Rybarczyk2022_Obs,Rybarczyk2023}. 
We present new CO absorption observations in three of these directions and new \hi{} observations in two of these directions. We also use archival CO absorption observations in three directions \citep[including one direction that we re-observed here for improved sensitivity;][]{Luo2020} and archival \hi{} absorption observations in three directions (\citealt{21SPONGE_2015,21SPONGE_2018}; \citealt{Rybarczyk2022_Obs} previously discussed the \hi{} and \hcop{} in these three directions).
The CO absorption observations trace CO at almost exactly the same angular scale as the \hcop{} observations, and with much greater CO column density sensitivity than observations of CO emission \citep[e.g.,][]{LisztLucas1998}, allowing us to place more reliable constraints on the column densities and abundances of CO.
In an effort to be consistent with the parlance used in the existing literature (which has focused largely on CO emission) while incorporating these new observations, we explicitly define different categories of gas based on the presence or absence of \hcop{} absorption, CO absorption, and CO emission (Table \ref{tab:categories}). In particular, we use ``\codarka'' to describe molecular gas (as revealed by \hcop{} absorption) with no associated CO emission \textit{or} absorption, while we use ``\codarkb'' to refer to molecular gas with associated CO absorption but no associated CO emission. We use ``CO-bright'' to describe gas with both CO absorption and CO emission.\footnote{In other works \citep[e.g.,][]{LisztPety2012}, a further distinction has been made to characterize CO-bright gas based on the observed brightness of CO relative to the expected brightness given a standard conversion to CO intensity $W_{\mathrm{CO}}=N(\htwo{})/(2\times10^{20}~[\persc{}~(\mathrm{K~\kms})^{-1}])$ \citep[e.g.,][and references therein]{Bolatto2013}. Since we are focused primarily on CO absorption here, we do not make such a distinction in this work, and instead adopt the broader CO-bright definition in Table \ref{tab:categories}. We also do not detect CO emission without absorption in this work, so do not make a category for such gas.}
Meanwhile, the \hi{} absorption observations allow us to constrain the multiphase properties of the atomic gas, also at the tiny angular scales probed by pencil-beam absorption (though the background source structure may be slightly different between the 3~mm molecular absorption and 21~cm atomic absorption data). 
Together, these complementary observations make it possible to characterize and compare the physical properties of the diffuse molecular, dense molecular, and cold atomic gas, on matched spatial scales and at unprecedented sensitivity.

The paper is structured as follows. In Section \ref{sec:observations}, we present the new and archival observations of \hcop{} and CO absorption from the Northern Extended Millimetre Array interferometer (NOEMA) and the Atacama Large Millimeter/submillimeter Array (ALMA) and \hi{} absorption from the Karl G. Jansky Very Large Array (VLA). In Section \ref{sec:methods}, we explain how \hcop{}, CO, and \hi{} absorption spectra are extracted from the interferometric observations, as well as how atomic and molecular column densities are determined from the absorption spectra. We then present the spectra in Section \ref{sec:results} and directly compare the \hcop{}, CO, and \hi{} properties in these directions. In Section \ref{sec:discussion}, we discuss the implications of these results in the context of observational and theoretical studies of molecule formation in the ISM. Finally, we present our conclusions in Section \ref{sec:conclusions}.

\section{Observations} \label{sec:observations}

\begin{deluxetable*}{c|c|c|c|c|c|c|c|c} \label{tab:observations}
\tablecaption{Sensitivities and flux densities for our sources}
\tablehead{
\colhead{Sightline} & \colhead{$\ell$} &  \colhead{$b$} & \colhead{$\sigma_{\tau_{\hcop{}}}$} & \colhead{$\sigma_{\tau_{\rm{CO}}}$} & \colhead{$\sigma_{\tau_{\rm{HI}}}$} & \colhead{$F_{89}$} & \colhead{$F_{115}$}  & \colhead{$E(B-V)$} \\
\colhead{} & \colhead{deg} &  \colhead{deg} & \colhead{} & \colhead{} & \colhead{} & \colhead{Jy} & \colhead{Jy} & \colhead{mag}
}
\startdata
    3C111 & 161.7 & -8.8 & 0.0012 & 0.0040 & 0.0023 & 1.15/2.25 & 1.56 & $1.117 \pm 0.056$ \\ 
    NRAO~530 & 12.0 & 10.8 & 0.0026 & 0.0067 & 0.0017 & 1.97 & 1.60 & $0.511 \pm 0.012$ \\
    J2023+335 & 73.1 & -2.4 & 0.0009 & 0.0044 & 0.0027 & 1.01 & 2.04 & $1.248 \pm 0.012$ \\
    3C454.3 & 86.1 & -38.2 & 0.0008 & 0.0024 & 0.0015 & 14.34 & 11.96 & $0.104 \pm 0.004$ \\
    3C120 & 190.4 & -27.4 &  0.0050 & 0.0127 & 0.0011 & 3.30 & 2.46  & $0.267 \pm 0.007$ \\
\enddata
\end{deluxetable*}

\subsection{New and archival molecular line observations with NOEMA} \label{sec:observations_with_NOEMA}
We have observed the $J$=(1--0) transition of CO ($115.2712~\mathrm{GHz}$) in absorption against the background radio continuum sources 3C111, NRAO 530, and J2023+335 (Table \ref{tab:observations}) with NOEMA (project W21AC). 
We placed a high-resolution chunk (62.5~kHz channel spacing) in the upper sideband to cover the CO line. While the upper sideband has somewhat poorer sensitivity than the lower sideband, the CO transition is near the highest frequencies observable in NOEMA Band 1. The bright sources 3C84, 3C273, 3C345, 3C454.3, and 2013+370 were used as bandpass calibrators\footnote{For NOEMA bandpass calibrations with mm absorption present, the relevant absorption is recognized by the pipeline and masked during bandpass calibration.}. The bandpass was stable across all our observations, and previous work (e.g., \citealt{Rybarczyk2022_Obs} and earlier work by, e.g., \citealt{Liszt2005,Liszt2014}, and references therein) has shown that optical depths $\lesssim10^{-3}$ can reliably be recovered with NOEMA (and previously the Plateau de Bure interferometer) using similar setups.
Targets were observed 
for $0.4$--$4.1~$hours.
All data reduction steps were done using the CLIC and MAPPING programs within the GILDAS software collection \citep{Pety2005,GILDAS2013}\footnote{https://www.iram.fr/IRAMFR/GILDAS/}.
We performed self-calibration using MAPPING's built-in tool, since our sources all had flux densities $\gtrsim1$~Jy (Table \ref{tab:observations}).

\hcop{} absorption has previously been observed with NOEMA toward the bright background sources 3C111, J2023+335, and NRAO530. 
\citet{Rybarczyk2023} presented \hcop{} observations toward J2023+335 and NRAO530. We use their \hcop{} absorption spectra here.
\citet{Rybarczyk2022_Obs} presented \hcop{} observations toward 3C111 (``3C111A'' in their work).
Here we have obtained additional, more-sensitive \hcop{} absorption observations in the direction of 3C111 (project W24BG) using an identical setup to that used by \citet{Rybarczyk2022_Obs}. We placed the \hcop{} $J=$(1--0) line ($89.1885~\mathrm{GHz}$) in a high resolution chunk (62.5~kHz channel spacing) in the lower sideband to ensure maximum sensitivity and observed 3C111 for 3.4~hours.
The final \hcop{} absorption spectrum toward 3C111 was calculated by combining the spectrum presented by \citet{Rybarczyk2022_Obs} with our new observations (weighted by the inverse-square of the noise in the \hcop{} absorption spectra).
As with CO, all reduction steps (including self-calibration) were carried out using the CLIC and MAPPING programs in GILDAS.

Table \ref{tab:observations} lists the coordinates, 89~GHz and 115~GH flux densities ($F_{89}$ and $F_{115}$; the difference between the flux densities at 89~GHz and 115~GHz are due both to the continuum spectral index as well as changes in the brightness of the background sources over time), and CO and \hcop{} optical depth sensitivities achieved in the direction of our background sources.
We present the CO and \hcop{} absorption spectra at $0.4~\kms{}$ velocity resolution in Figure \ref{fig:spectra} and discuss the spectra in Section \ref{sec:results}.

\subsection{Archival molecular line observations with ALMA} \label{sec:observations_with_ALMA}
\citet{Luo2020} observed the $J$=(1--0) transition of CO in absorption using the Atacama Large Millimeter/submillimeter Array (ALMA) toward 3C120, 3C454.3, and NRAO530 (2015.1.00503.S). Meanwhile, both
\citet{Luo2020} (2015.1.00503.S) and \citet{Rybarczyk2022_Obs} (2018.1.00585.S) observed the $J$=(1--0) transition of \hcop{} in absorption with ALMA toward 3C120 and 3C454.3.
These ALMA Band 3 observations are discussed in detail by \citet{Luo2020} and \citet{Rybarczyk2022_Obs}. 

Here we use the CO absorption spectra obtained by \citet{Luo2020} toward 3C120 and 3C454.3. 
For NRAO530, we combine the \citet{Luo2020} CO absorption spectra with the new CO absorption spectra observed here (Section \ref{sec:observations_with_NOEMA}), taking a weighted mean (weighted by the inverse square of the noise in $e^{-\tau}$).
Similarly, for 3C120 and 3C454.3, we combine the \hcop{} absorption spectra from \citet{Luo2020} and \citet{Rybarczyk2022_Obs}, again taking a weighted mean. 
We smooth all ALMA absorption spectra to $0.4~\kms{}$ velocity resolution.
We list the noise levels in the final CO and \hcop{} spectra in these directions in Table \ref{tab:observations}.
The ALMA CO and \hcop{} absorption spectra are shown in Figure \ref{fig:spectra}.

\subsection{Archival CO emission observations from the CfA 1.2~m telescope} \label{sec:Dame}
While we focus on CO absorption in this work, we also show archival CO emission spectra from the 1.2 m Millimeter-Wave Telescope at the Center for
Astrophysics \citep{DameThaddeus2001,Dame2022} toward all five of our sightlines. These spectra have typical noise $\lesssim0.2~\mathrm{K}$ at $0.65~\kms{}$ velocity resolution. The angular resolution achieved in these directions was $\leq0.25^\circ$. The CO emission spectra are shown in Figure \ref{fig:spectra}.

\subsection{New and archival \hi{} absorption and emission observations} \label{subsec:HI_obs}
We use observations of \hi{} absorption and emission at $1420~\mathrm{MHz}$ in the direction of all five background targets to constrain the properties of the atomic gas.
Previously, \citet{21SPONGE_2015,21SPONGE_2018} observed \hi{} absorption with the Karl G. Jansky Very Large Array (VLA) toward 3C111 (``3C111A'' in their work), 3C120, and 3C454.3 as part of the 21-SPONGE project. They achieved an \hi{} optical depth sensitivity $(0.9$--$1.7)\times10^{-3}$ in these directions at $\sim0.4~\kms{}$ velocity resolution. They also extracted emission spectra in these directions using the  Galactic Arecibo L-Band Feed Array \hi{} survey \citep[GALFA-\hi{};][]{GALFA2011,GALFA2018}. We use the 21-SPONGE spectra for 3C111, 3C120, and 3C454.3 in this work.

We have further observed \hi{} in absorption with the VLA toward J2023+335 and NRAO530 (project 24A-088) with a nearly identical observing setup and reduction process to that used by the 21-SPONGE project (see \citealt{21SPONGE_2015} for details; we briefly summarize the observations and reduction here). 
NRAO530 was observed for 0.7~hours. J2023+335 was observed for $6.4~$hours in total, spread across two observations lasting $3.1~$hours and $3.3~$hours, respectively. Both sources were observed in the VLA's C configuration; we excluded baselines shorter than $300~\mathrm{m}$ to avoid possible contamination from extended or large-scale \hi{} emission.
We used three 1.0~MHz bands with 1.95~kHz channel spacing --- one band was centered on the \hi{} line, and the other two bands were offset by $\pm1.5~$MHz from the \hi{} line. The offset bands were used to perform bandpass calibration via frequency switching to avoid contamination by Galactic \hi{} absorption. 3C286 was used for bandpass calibration for NRAO530 and J2052+3635 was used for bandpass calibration in the direction of J2023+335.
The final absorption spectra are shown in Figure \ref{fig:spectra} and discussed in Section \ref{sec:results}.
\hi{} optical depth sensitivities at $0.4~\kms{}$ velocity resolution are listed in Table \ref{tab:observations}.

We also extracted the \hi{} emission profiles in these directions using data from the GALFA-\hi{} survey \citet[][]{GALFA2011,GALFA2018} for J2023+335 and from the Parkes Galactic All-Sky Survey \citep[GASS;][]{GASS2009} for NRAO530 using the approach introduced by \citet{HT03}. Essentially, to avoid contamination of the emission spectra by the continuum sources, we extracted the emission spectra using an interpolation technique with nearby pixels (16 pixels separated by 1 to $\sqrt{2}$ times the half-power beam width; see \citealt{HT03} for a complete description of the technique). This approach is common in previous \hi{} emission-absorption studies \citep[e.g.,][]{HT03,SS2014,21SPONGE_2015}.

\subsection{Reddening measurements} \label{subsec:reddening}
To provide further context for our observations, in Table \ref{tab:observations} we also report $E(B-V)$ measured in each direction by \citet{Green2019}. The 3D reddening measurements from \citet{Green2019} are based on stellar photometry from Pan-STARRS 1 and the Two Micron All Sky Survey, as well as parallaxes from \textit{Gaia}. Here we report only $E(B-V)$ for the full line of sight. We convert Bayestar reddening $E$ to $E(B-V)$ using the standard $E(B-V)=0.883 E$ \citep{Schlafly2011,Green2019}.

\section{Methods} \label{sec:methods}
In this work, we use absorption observations against background continuum sources, bright at both 1420~MHz and $\sim100~\mathrm{GHz}$, to determine the physical properties of the atomic and molecular gas in five different directions. We use a standard approach, described here, to extract the absorption spectra and constrain the column densities.

If the observed flux density at frequency $\nu$ of a background continuum source is $S(\nu)$ and the continuum flux density (i.e., the flux density not modified by absorption) is $S_0(\nu)$, then the absorption spectrum is defined as
\begin{equation} \label{eq:emtau}
    e^{-\tau(\nu)} = \frac{S(\nu)}{S_0(\nu)}.
\end{equation}
To measure $S_0(\nu)$ in our data, we fit a first-order polynomial to $S(\nu)$ for channels with no absorption. 
The first-order polynomial removes potential structure in the continuum spectrum introduced by the spectral index (this approach was previously used by \citealt{Rybarczyk2022_Obs} for the \hcop{} spectra in the direction of 3C111, 3C120, and 3C454.3).
We note that Equation \ref{eq:emtau} assumes that either the continuum brightness temperature is much larger than the excitation temperature of the relevant transition, or that  emission in that transition is spatially extended such that it will be filtered out by the interferometer, or both. 

\begin{figure*}[]
    \centering
    \includegraphics[width=\linewidth]{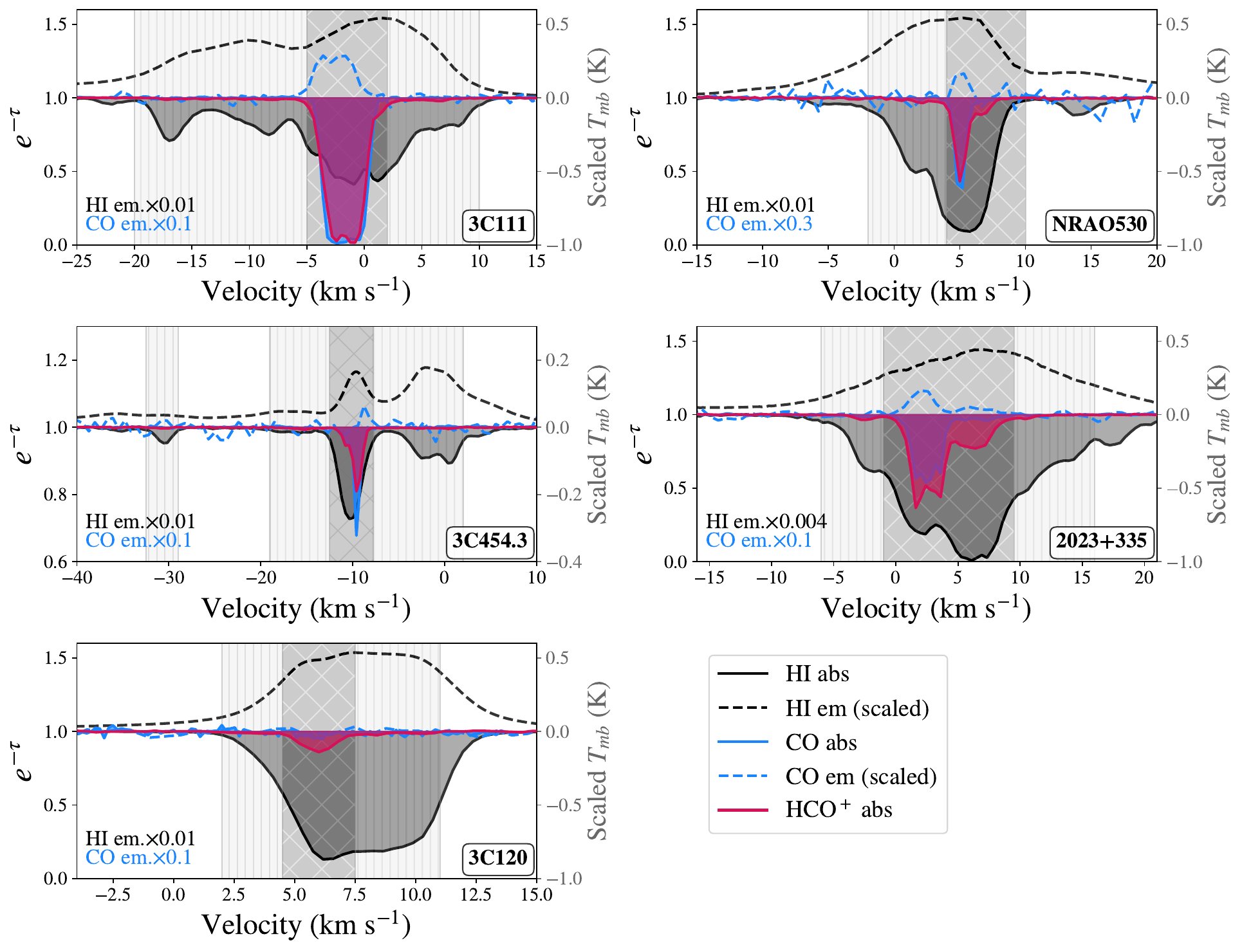}
    \caption{The absorption spectra, $e^{-\tau}$, of \hi{} (filled black), \hcop{} (filled pink), and CO (filled blue) in the direction of the five background sources listed in Table \ref{tab:observations}. Velocities where narrow \hcop{} absorption is observed are highlighted in cross-hatched dark gray (see Figure \ref{fig:HI_CO_HCOp_zoomin}, which highlights the broad, weak spectral features, difficult to see in these plots). Velocities where broad \hcop{} absorption is observed are highlighted in vertically hatched light gray. \hi{} emission (dashed black line) and CO emission (dashed blue line) are also shown for context; the right-hand $y$-axis shows the brightness temperature for \hi{} and CO emission, scaled by the factors  indicated in the bottom left corner of each plot.}
    \label{fig:spectra}
\end{figure*}

These criteria should be met for all our observations, as the molecular lines have very low excitation temperatures (see discussion below) and we deliberately remove short baselines from the \hi{} observations to prevent contamination from resolved emission. 
Moreover, if we remove the shortest baselines (we tested for baselines $<100~$m and $<200~$m)  from our NOEMA and ALMA CO and \hcop{} observations, we find statistically insignificant changes in the channel-by-channel optical depths or integrated optical depth in all but one case. 
The only spectrum for which we detect significant changes in the optical depth when removing short baselines is the CO spectrum in the direction of 3C111, where we detect up to $\sim5\sigma$ changes in two channels. However, these are channels in the range where CO absorption is saturated, so the optical depth is already unconstrained (see discussion below). This is largely consistent with expectations --- if we assume a CO excitation temperature of $4~\mathrm{K}$ in the diffuse ISM \citep{Luo2020}, and a brightness temperature of CO emission of $\sim1~\mathrm{K}$ at small (sub-arcminute) scales (\citealt{LisztLucas1998}; see detailed maps in the directions of 3C454.3 and NRAO~530 in \citealt{LisztPety2012}), then in most cases, the potential CO emission, if it has not been filtered out by interferometer, would contribute no more than a few percent of the measured background brightness temperature at our current angular resolution. 
Thus, they have a negligible effect on the measured optical depths. 
The CO emission is significantly stronger in the direction of 3C111, though ($\sim4~\mathrm{K}$, \citealt{LisztLucas1998}; see also the lower-resolution CO emission spectra in Figure \ref{fig:spectra}). The same arguments hold for \hcop{}, but \hcop{} emission is often significantly weaker than CO emission \citep{LucasLiszt1996}. And again, the potential contamination from emission will only align with denser regions where the \hcop{} absorption is particularly strong. In summary, contamination from emission will have a minimal impact on our measured optical depths in all but one case (and in that one case --- the CO spectrum towards 3C111 --- we do not report the optical depth at the contaminated velocity due to line saturation). 
In order to preserve the best noise possible, we therefore do include all baselines used by ALMA and NOEMA for CO, noting that an additional uncertainty of $\sim\mathrm{few}\%$ may result in regions of strong CO absorption.

Once we have extracted the absorption spectrum for a particular transition from upper state $u$ to lower state $l$, at frequency $\nu_0$, we can solve for the column density 
\begin{equation} \label{eq:N}
    N = Q(T_{\mathrm{ex}})\frac{8\pi\nu^3}{c^3}\frac{1}{g_u A_{ul}}\Big[1-\exp\Big(-\frac{h\nu_0}{kT_{\mathrm{ex}}}\Big)\Big]^{-1} \int\tau(v) dv.
\end{equation}
Here, $g_u$ is the degeneracy of the upper energy state, $A_{ul}$ is the Einstein $A$ coefficient for the transition, $T_{\mathrm{ex}}$ is the excitation temperature, and $Q(T_{\mathrm{ex}})$ is the partition function. 
We have also expressed the optical depth in terms of velocity $v$ rather than frequency in Equation \ref{eq:N}.

\begin{deluxetable}{c|c|c} \label{tab:tau_to_N}
\tablecaption{Conversions from $\int \tau dv$ to column density}
\tablehead{
\colhead{\texc{}} & \colhead{$N/\int\tau_{\mathrm{CO}} dv$} &  \colhead{$N/\int\tau_{\hcop{}} dv$} \\
\colhead{K} & \colhead{$10^{15}\persc{}/\kms{}$} &  \colhead{$10^{12}\persc{}/\kms{}$} 
}
\startdata
    $\tcmb{}=2.725$ & $1.07$ & $1.09$ \\
    $4.1$ & $1.67$ & $1.83$ \\
    $5$ & $2.16$ & $2.43$ \\
    $10$ & $6.18$ & $7.48$\\
\enddata
\end{deluxetable}

Without absorption observations of higher-level rotational transitions, we cannot measure the excitation temperature of \hcop{} or CO directly.
Previous measurements of the \hcop{} excitation temperature in the diffuse/translucent ISM have all been consistent with $T_{\mathrm{ex}} \approx T_{\rm CMB}=2.725~\mathrm{K}$ \citep[$2.7$--$3.0~\rm{K}$;][]{Godard2010,Luo2020}. \citet{Luo2020} specifically derived an excitation temperature $\texc=2.7~\mathrm{K}$ for \hcop{} in the direction of 3C454.3. 
Meanwhile, \citet{Goldsmith2013} measured CO excitation temperatures $2.7$--$13.6~\rm{K}$ (with $63/64$ measured excitation temperatures $\leq6.0~\rm{K}$) from UV measurements in the direction of stars in the solar neighborhood, probing diffuse gas. \citet{Luo2020} measured the $J$=(1--0) and $J$=(2--1) transitions of CO in the direction of 3C454.3, deriving a CO excitation temperature $\texc=4.1~\rm{K}$. 
More generally, {\sc radex} modeling of the CO $J$=(1--0) and $J$=(2--1) transitions toward four diffuse sightlines by \citet{Luo2023} yields excitation temperatures of $\sim$3--4~K. 
These sub-thermal excitation temperatures should be appropriate for the majority of the CO detections in our sample. 

In Table \ref{tab:tau_to_N}, we give the conversion between integrated optical depth and column density (Equation \ref{eq:N}) for CO and \hcop{} for a range of values of \texc.
Throughout the rest of this work, we assume $\texc=\tcmb$ for \hcop{} for all directions. 
Because the CO excitation temperature is less tightly constrained in the diffuse ISM, we consider a range of possible CO excitation temperatures moving forward.

Meanwhile, for \hi{}, the excitation temperature (generally referred to as the ``spin temperature,'' $T_s$) is much higher than \tcmb{}. 
In this work, we calculate spin temperatures using an isothermal approximation, which assigns a single spin temperature to the atomic gas in each velocity channel (essentially ignoring the blending of different components that may have differing $T_{\rm s}$). This is equal to $T_{\rm s}(v)=T_{\rm B}(v)(1-e^{-\tau(v)})^{-1}$. 
The \hi{} column densities are then calculated as $N(\hi{}) = 1.823\times10^{18} \times \int  T_{\rm s}(v) \tau(v) dv$.
In the future, we will consider \hi{} column densities for discrete \hi{} clouds along each line of sight (see Section \ref{subsubsec:discussion_comparison_to_OH}), but the isothermal approximation to measuring the integrated column density agrees reasonably well with this more sophisticated approach in most cases \citep[e.g.,][]{21SPONGE_2018}.

\section{Results} \label{sec:results}
\begin{figure*}
    \centering
    \includegraphics[width=\linewidth]{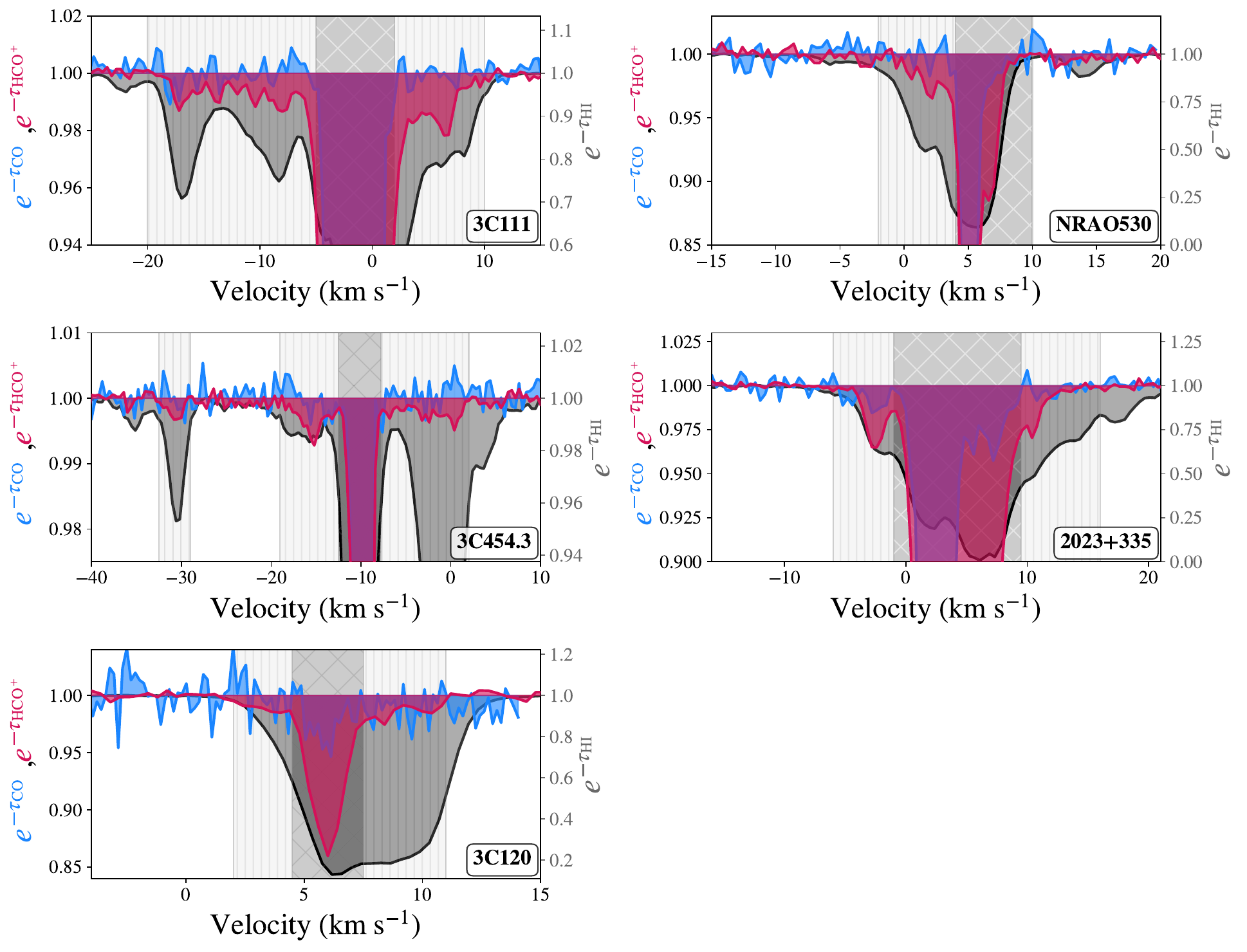}
    \caption{Same as Figure \ref{fig:spectra}, but with restricted $y$-axis ranges to highlight the broad \hcop{} absorption component. The left $y$-axis shows $e^{-\tau}$ for CO and \hcop{}, while the right axis shows $e^{-\tau}$ for \hi{}. We do not show the \hi{} and CO emission here.}
    \label{fig:HI_CO_HCOp_zoomin}
\end{figure*}

We detect absorption from CO, \hcop{}, and \hi{} toward all five background radio continuum sources.
In Figures \ref{fig:spectra} and \ref{fig:HI_CO_HCOp_zoomin}, we show the absorption spectra, $e^{-\tau(v)}$, for CO (blue), \hcop{} (pink), and \hi{} (black) in all five directions. 
Figure \ref{fig:spectra} shows the entire spectra, as well as the \hi{} and CO emission spectra in the same directions (dotted lines) for context, while Figure \ref{fig:HI_CO_HCOp_zoomin} focuses on the  shallow, broad \hcop{} absorption discussed by \citet{LisztLucas2000} and \citet{Rybarczyk2022_Obs,Rybarczyk2023}.

In this work, we are focused primarily on understanding the origin and physical properties of the gas traced by broad \hcop{} absorption. 
We must therefore explicitly define what is meant by broad versus narrow.
In previous work, ``narrow'' has referred to spectral intervals where the strongest \hcop{} absorption is observed (on a sightline-by-sightline basis), confined to a relatively narrow range of velocities ($\sim\mathrm{few}~\kms{}$) and comprising one or a few narrow peaks. Conversely, ``broad'' has referred to spectral intervals where the \hcop{} absorption is notably weaker, typically extending several \kms{} to one or both sides of the narrow absorption   \citep{LisztLucas2000,Rybarczyk2022_Obs}. 
\citealt{Rybarczyk2022_Obs} earlier defined the ``broad'' and ``narrow'' regions for 3C111, 3C120, and 3C454.3 based solely on the \hcop{} absorption profiles; for consistency, we adopt their bounds for these sightlines. For our other two sightlines, we adopt a similar approach to defining broad and narrow absorption.
Namely, we use ``narrow'' (on a sightline-by-sightline basis) to refer to spectral intervals where the strongest \hcop{} absorption is observed (typically, $\tau_{\mathrm{HCO^+}}\gtrsim0.1$), confined to a relatively narrow range of velocities ($\sim\mathrm{few}~\kms{}$) comprising one or a few narrow peaks.
Regions of narrow absorption are highlighted in cross-hatched dark gray in Figures \ref{fig:spectra} and \ref{fig:HI_CO_HCOp_zoomin}.
Meanwhile, we use ``broad'' to refer to spectral intervals where the \hcop{} absorption is weaker and extends several \kms{} beyond where the narrow absorption lines are observed. 
The broad absorption does not necessarily comprise absorbing components with particularly broad linewidths, but contributes to the overall breadth of the \hcop{} absorption in a given direction. 
Regions of narrow absorption are highlighted in cross-hatched dark gray in Figures \ref{fig:spectra} and \ref{fig:HI_CO_HCOp_zoomin}, whereas regions of broad absorption are highlighted in  vertically-hatched light gray.
Median signal-to-noise values for the broad spectral regions are $\sim2.5$--7.

We note that while we adopt this categorization for consistency with the existing literature \citep[][]{LisztLucas2000,Rybarczyk2022_Obs}, we do not mean to automatically imply physical or chemical differences. 
Later we discuss how both the broad and narrow absorption features fit into our understanding of the diffuse ISM.

\begin{deluxetable*}{|cccrrr|} \label{tab:int_tau_CO_HCOp}
\tablecaption{The integrated CO and \hcop{} optical depths and their ratios for each region outlined in Figure \ref{fig:HI_CO_HCOp_zoomin}.}
\tablehead{
\colhead{Sightline} & \colhead{Type} & \colhead{Velocity interval} & \colhead{$\int\tau_{\hcop{}} dv$} &  \colhead{$\int\tau_{\mathrm{CO}} dv$} & \colhead{$\frac{\int\tau_{\hcop{}} dv}{\int\tau_{\mathrm{CO}} dv}$} \\
\colhead{} & \colhead{} & \colhead{\kms{}} & \colhead{\kms{}} & \colhead{\kms{}} & \colhead{} 
}
\startdata
3C111 & none & $[-40.0,-20.0)$ & $<0.010$ & $<0.034$ & $\cdot\cdot\cdot$ \\
      & broad & $[-20.0,-5.0)$ & $0.118 \pm 0.003$ & $<0.029$ & $>4.073$ \\
      & narrow & $[-5.0,2.0)$ & $11.108$\tablenotemark{a} & $13.459$\tablenotemark{a} & $0.825$\tablenotemark{a} \\
      & broad & $[2.0,10.0)$ & $0.129 \pm 0.002$ & $<0.021$ & $>6.045$ \\
      & none & $[10.0,40.0)$ & $<0.012$ & $<0.041$ & $\cdot\cdot\cdot$ \\
\hline
NRAO530 & none & $[-20.0,-2.0)$ & $<0.021$ & $<0.054$ & $\cdot\cdot\cdot$ \\
        & broad & $[-2.0,4.0)$ & $0.102 \pm 0.004$ & $<0.030$ & $>3.366$ \\
        & narrow & $[4.0,10.0)$ & $0.985 \pm 0.005$ & $0.941 \pm 0.010$ &  $1.047 \pm 0.013$ \\
        & none & $[10.0,20.0)$ & $0.019 \pm 0.005$ & $<0.040$ & $>0.462$ \\
\hline
J2023+335 & none & $[-40.0,-6.0)$ & $<0.010$ & $<0.049$ & $\cdot\cdot\cdot$ \\
         & broad & $[-6.0,-1.0)$ & $0.074 \pm 0.001$ & $0.039 \pm 0.006$ &  $1.915 \pm 0.307$ \\
         & narrow & $[-1.0,9.5)$ & $3.319 \pm 0.003$ & $1.646 \pm 0.009$ &  $2.016 \pm 0.011$ \\
         & broad & $[9.5,16.0)$ & $0.058 \pm 0.001$ & $<0.021$ & $>2.714$ \\
         & none & $[16.0,40.0)$ & $<0.009$ & $<0.041$ & $\cdot\cdot\cdot$ \\
\hline
3C454.3 & none & $[-40.0,-32.5)$ & $<0.004$ & $<0.008$ & $\cdot\cdot\cdot$ \\
        & broad & $[-32.5,-29.0)$ & $0.006 \pm 0.001$ & $<0.006$ & $>1.126$ \\
        & none & $[-29.0,-19.0)$ & $0.006 \pm 0.002$ & $<0.010$ & $>0.655$ \\
        & broad & $[-19.0,-12.5)$ & $0.020 \pm 0.001$ & $<0.008$ & $>2.640$ \\
        & narrow & $[-12.5,-7.75)$ & $0.282 \pm 0.001$ & $0.345 \pm 0.002$ &  $0.817 \pm 0.006$ \\
        & broad & $[-7.75,2.0)$ & $0.022 \pm 0.002$ & $<0.009$ & $>2.359$ \\
        & none & $[2.0,10.0)$ & $<0.004$ & $<0.009$ & $\cdot\cdot\cdot$ \\
\hline
3C120 & none & $[-10.0,2.0)$ & $<0.033$ & $<0.043$ & $\cdot\cdot\cdot$ \\
      & broad & $[2.0,4.5)$ & $0.029 \pm 0.005$ & $<0.025$ & $>1.169$ \\
      & narrow & $[4.5,7.5)$ & $0.247 \pm 0.005$ & $0.040 \pm 0.009$ &  $6.098 \pm 1.427$ \\
      & broad & $[7.5,11.0)$ & $0.055 \pm 0.006$ & $<0.031$ & $>1.791$ \\
      & none & $[11.0,15.0)$ & $<0.019$ & $<0.028$ & $\cdot\cdot\cdot$ \\
\enddata
\tablenotetext{a}{Because of saturation for both CO and \hcop{} in the direction of 3C111, the integrated optical depths are both lower limits, so the ratio is not constrained.}
\end{deluxetable*}

\subsection{Integrated \hcop{} and CO optical depths}
\label{subsec:integrated_HCOp_and_CO}

In Table \ref{tab:int_tau_CO_HCOp}, we list the velocity bounds for each of the broad and narrow regions in Figures \ref{fig:spectra} and \ref{fig:HI_CO_HCOp_zoomin}, as well as their integrated CO and \hcop{} optical depths.
Narrow \hcop{} absorption is detected in all directions. The narrow \hcop{} absorption is associated with the strongest \hi\ absorption in each direction and is coincident with narrow CO absorption in all cases. 
The broad \hcop{} component extends across nearly all velocities where \hi{} is detected in absorption, rather than just the optically thickest \hi{}  associated with the strong, narrow molecular features. 
Toward J2023+335, we detect some CO associated with broad \hcop{} absorption, but most of the broad \hcop{} absorption is \codarka\ at the sensitivity achieved here (Table \ref{tab:observations}).

\begin{figure}
    \centering
    \includegraphics[width=\linewidth]{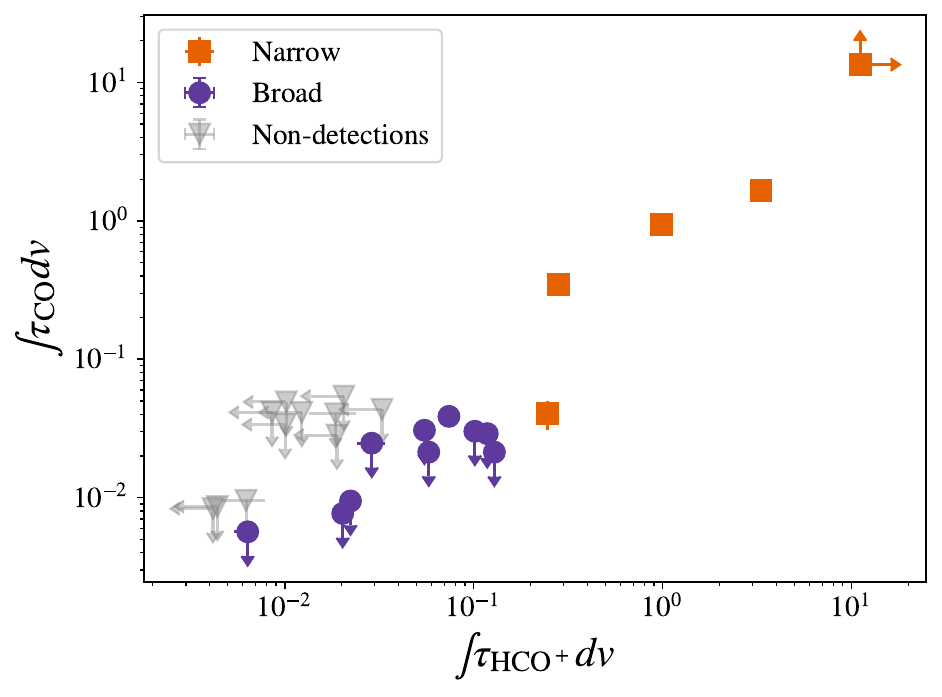}
    \caption{The integrated CO optical depth versus the integrated \hcop{} optical depth for different velocity intervals (listed in Table \ref{tab:int_tau_CO_HCOp}). Results for regions with broad \hcop{} absorption are shown as purple circles. Results for regions with narrow \hcop{} absorption are shown as orange squares. For non-detections, $3\sigma$ upper limits are shown as semi-transparent gray triangles. For points without clear error bars, the uncertainties are smaller than the sizes of the points.}
    \label{fig:inttauCO_v_inttauHCOp}
\end{figure}

In Figure \ref{fig:inttauCO_v_inttauHCOp}, we show the integrated CO optical depth versus the integrated \hcop{} optical depth for all regions listed in Table \ref{tab:int_tau_CO_HCOp}. Regions with narrow \hcop{} absorption are shown as orange squares and regions with broad \hcop{} absorption are shown as purple circles. There is a clear separation in this space, with the narrow absorption having systematically higher $\int\tau_{\mathrm{CO}}dv$ and $\int\tau_{\hcop}dv$ (in the latter case, this is expected from the definition of narrow absorption, as discussed above). 
CO is detected in all five regions with narrow \hcop{} absorption, but in only one of the 10 regions with broad \hcop{} absorption (J2023+335).
It is notable here that the narrow absorption toward 3C120 (the orange point with the lowest $\int\tau_{\mathrm{CO}}dv$) lies at the interface of the broad and narrow regimes on this plot,
perhaps suggesting that it is tracing a more diffuse environment than the other narrow components.

From Figure \ref{fig:inttauCO_v_inttauHCOp}, we find an approximate threshold of $\int\tau_{\hcop}\sim0.2~\kms{}$ below which most of the diffuse molecular gas is \codarka.  

If we assume $N(\hcop{})/N(\htwo{})=3\times10^{-9}$ \citep{LisztGerin2023a} and an \hcop{} excitation temperature 2.725~K (which is well justified; see Section \ref{sec:methods}), then this implies that the diffuse molecular gas in our sample is mostly \codarka\ below a threshold of $N(\htwo{})\sim7\times10^{19}~\persc{}$.
The only exception is the detection of CO toward the broad component of J2023+335, associated with an \hcop{} column density $\sim8\times10^{10}~\persc{}$, implying an \htwo{} column density $\sim3\times10^{19}~\persc{}$.

On a sightline-by-sightline basis, we estimate that the fraction of the molecular hydrogen that is \codarka is between 3\% and 20\% (here taken as the fraction of the \hcop{} column density in channels with no CO absorption at a level of $3\sigma$). Naturally, the \codarka\ fraction will vary with environment; as expected, we find the highest \codarka\ fractions towards sightlines with lower molecular column densities. 

\subsection{The CO abundance}
\label{sec:CO_abundance}

\begin{figure}
    \centering
    \includegraphics[width=\linewidth]{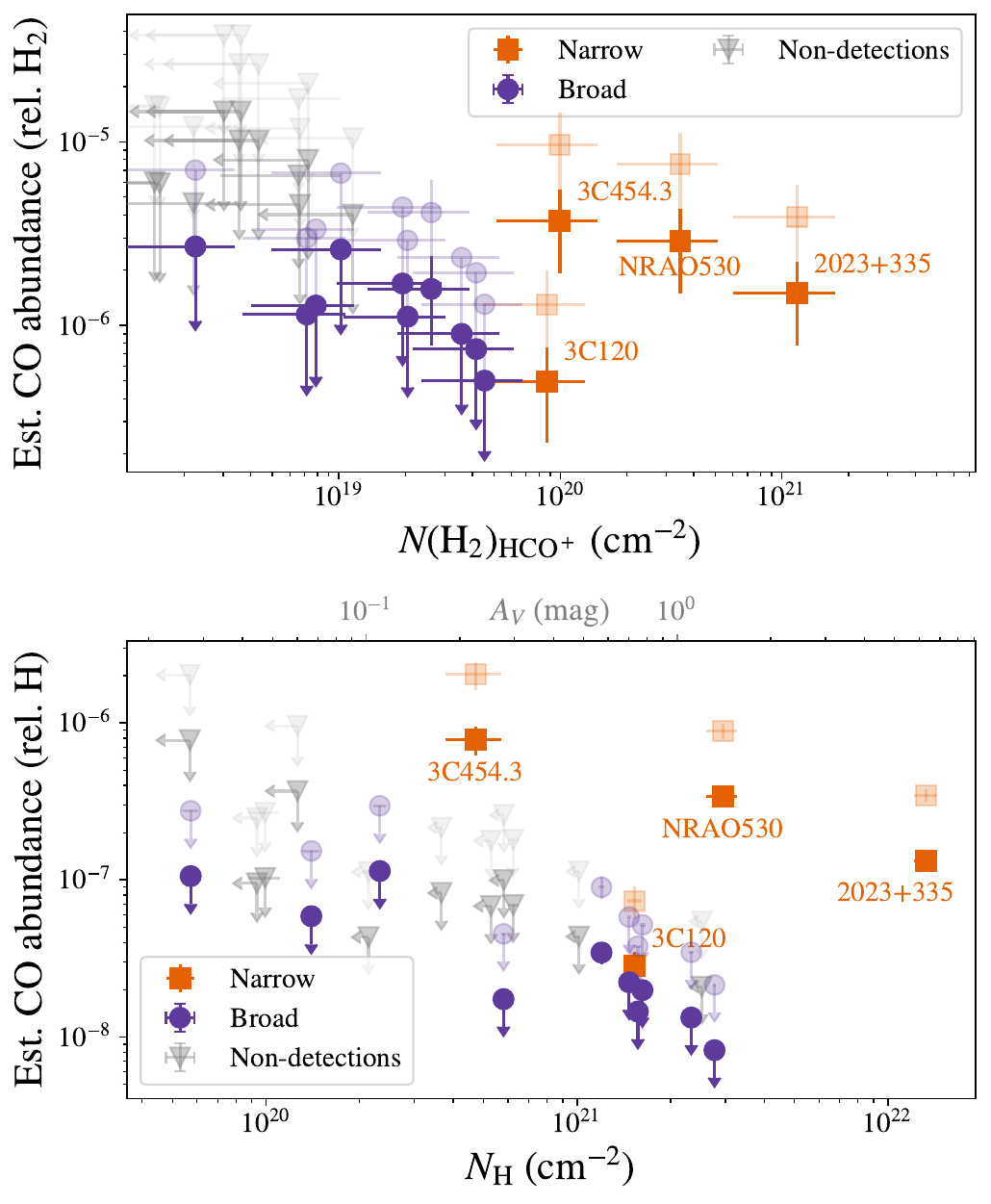}
    \caption{\textit{Top:} the implied CO abundance relative to \htwo{} versus the \htwo{} column density (derived from the \hcop{} column density) for all regions listed in Table \ref{tab:int_tau_CO_HCOp}. \textit{Bottom:} the implied CO abundance relative to the total hydrogen column density versus the total column density, $N_{\mathrm{H}}$ (derived from the \hcop{} column density and the \hi{} column density) for all regions listed in Table \ref{tab:int_tau_CO_HCOp}. 
    We also show $A_V$ in the bottom panel, where we convert to $A_V$ from $N_{\mathrm{H}}$ assuming $A_V= N_{\mathrm{H}}/2.1\times10^{21} ~\mathrm{mag}/\persc{}$ \citep{GuverOzel2009,Zhu2017}.
    In both plots, the results for regions with narrow absorption are shown with orange squares (labeled for each sightline) and the results for regions with broad absorption are shown with purple circles. Results for regions with neither \hcop{} nor CO detections are shown as gray triangles. The CO excitation temperature is unknown; solid points show results for a CO excitation temperature of $2.725~\mathrm{K}$ and semi-transparent points show results for a CO excitation temperature of $6~\mathrm{K}$.}
    \label{fig:implied_CO_abundance}
\end{figure}

In Figure \ref{fig:implied_CO_abundance}, we show 
the implied CO abundance relative to \htwo{} (top panel) and relative to all hydrogen nuclei (\hi+\htwo{}; bottom panel). Here,
$N(\rm \htwo{})$ is derived directly from $N(\rm\hcop{})$ using the assumed abundance of $3\times10^{-9}$, and $N({\rm \hi{}})$ is derived from emission and absorption as described in Section \ref{sec:methods}. 
We note that results for the narrow absorption toward 3C111 are not included --- because both CO and \hcop{} are saturated, the CO abundance is unconstrained.
Opaque points show the results for a CO excitation temperature of $T_{\mathrm{CMB}}$; semi-transparent points show the results for a CO excitation temperature of $6~\mathrm{K}$. Based on previous observations at similar column densities in the diffuse ISM \citep[e.g.,][]{Goldsmith2013,Luo2020,Luo2023}, true values likely lie between these two limits (see Section \ref{sec:methods}).

The implied CO abundances relative to \htwo{} are a $\mathrm{few}\times10^{-7}$ to a $\mathrm{few}\times10^{-6}$ for regions traced by narrow \hcop{} absorption and $\lesssim\mathrm{a\ few}\times10^{-6}$ (upper limits) for regions traced by broad \hcop{} absorption.
Similarly, the implied CO abundances relative to all hydrogen nuclei are $\mathrm{few}\times10^{-8}$ to $\mathrm{few}\times10^{-6}$ for regions traced by narrow \hcop{} absorption and $\lesssim\mathrm{few}\times10^{-7}$ (upper limits) for regions traced by broad \hcop{} absorption. 
As in Figure \ref{fig:inttauCO_v_inttauHCOp}, we find that the 
narrow absorption toward 3C120 (the orange point with the lowest abundance --- in both panels --- in Figure \ref{fig:implied_CO_abundance}) falls in a similar area of parameter space to the 
broad absorption in other directions. 
More generally, our measurements are not sufficient to conclude whether the CO abundance relative to \htwo{} is systematically lower for gas traced by broad absorption. 
We note that the abundances relative to all hydrogen nuclei (bottom panel of Figure \ref{fig:implied_CO_abundance}) appear systematically higher for regions traced by narrow absorption (excluding the direction of 3C120), but this is expected given the similar abundances (or limits) relative to \htwo{} (top panel of Figure \ref{fig:implied_CO_abundance}), combined with the lower molecular fractions for gas traced by broad absorption (see Section \ref{sec:hi_associated_with_molecules} below).

\subsection{Comparing \hi{} and diffuse molecular gas properties}
\label{sec:hi_associated_with_molecules}

\subsubsection{The CNM fraction and molecular gas fraction}
\label{subsubsec:CNM_fraction}
\begin{figure}
    \centering
    \includegraphics[width=\linewidth]{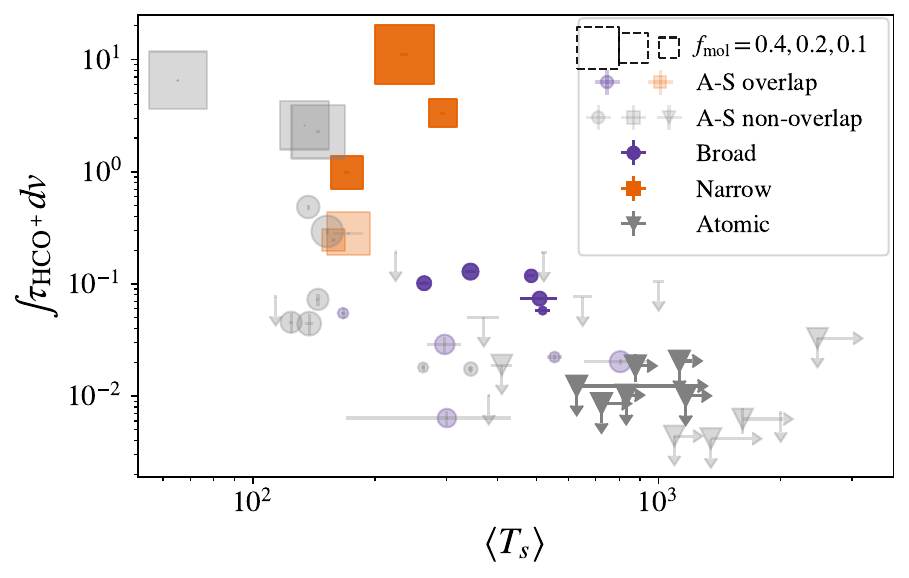}
    \caption{The integrated \hcop{} optical depth versus $\langle \ts \rangle$ (Equation \ref{eq:int_ts}). Results for regions with narrow \hcop{} absorption are shown as squares. Results for regions with broad \hcop{} absorption are shown as circles. Results for regions where only atomic gas is detected in absorption are shown as triangles. New results from this work are plotted in opaque orange (for narrow absorption regions), purple (for broad absorption regions) and solid gray (for regions with no molecular absorption) markers. Data from the ALMA-SPONGE project \citep{Rybarczyk2022_Obs} toward the sightlines discussed here are shown in the same color scheme with semi-transparent markers (labeled ``A-S overlap''). Data from \citet{Rybarczyk2022_Obs} in other directions are shown in semi-transparent gray (labeled ``A-S non-overlap''). Markers are sized according the the molecular fraction, $f_{\mathrm{mol}}$ (the legend shows marker sizes for $f_{\mathrm{mol}}=0.4$, $0.2$, and $0.1$ for reference).}
    \label{fig:inttauHCOp_v_intTs}
\end{figure}

\citet{Rybarczyk2022_Obs} previously showed that that the \hi{} associated with broad \hcop{} absorption had a lower CNM fraction than that associated with narrow \hcop{} absorption. Here we perform a similar analysis for our current sample, by comparing the integrated \hcop{} optical depth and the \hi\ optical-depth-weighted mean spin temperature,
\begin{equation} \label{eq:int_ts}
    \langle \ts \rangle = \frac{\int \tau(v)T_\mathrm{B}(v)/(1-e^{-\tau(v)})dv}{\int\tau(v)dv}.
\end{equation}
This quantity is not, in general, the actual physical temperature of the gas, as there may be multiple structures with different temperatures blended in velocity space. Instead, $\langle \ts \rangle$ is a proxy for the fraction of \hi{} in the CNM, being inversely proportional to $f_{\mathrm{CNM}}=N(\hi{}_{\mathrm{CNM}})/N(\hi{})$  \citep{Kim2014}. 
Figure \ref{fig:inttauHCOp_v_intTs} shows the integrated \hcop{} optical depth versus $\langle \ts \rangle$ for regions with narrow \hcop{} absorption (orange squares), broad \hcop{} absorption (purple circles), and with no \hcop{} absorption (triangles). Results from \citet{Rybarczyk2022_Obs} are also overplotted.\footnote{\citet{Rybarczyk2022_Obs} considered HI and \hcop{} toward 3C111, 3C120, and 3C454.3 \citep[including data from][]{Luo2020}, as well as four additional sightlines. In Figure \ref{fig:inttauHCOp_v_intTs}, the results for 3C120 and 3C454.3 are plotted with semi-transparent colored markers to indicate the overlap with this work, whereas the results for 3C111 are plotted as opaque points, since we re-calculated $\int\tau_{\hcop}dv$ for the new, more sensitive spectra presented here (see Section \ref{sec:observations_with_NOEMA}).
We also note that \citet{Rybarczyk2022_Obs} did not observe CO absorption in any direction.}
Points are sized according to the molecular fraction,
$f_{\mathrm{mol}}=2N(\htwo{})/[N(\hi{})+2N(\htwo{})]$.

\begin{figure}
    \centering
    \includegraphics[width=\linewidth]{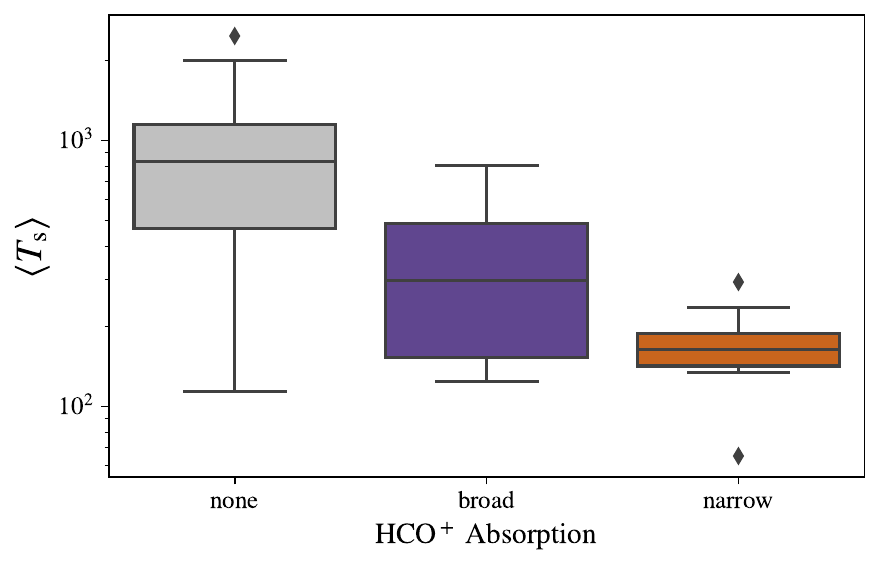}
    \caption{Box plot of $\langle \ts \rangle$ (Equation \ref{eq:int_ts}) for regions with narrow \hcop{} absorption (orange, rightmost), broad \hcop{} absorption (purple, middle), and no \hcop{} absorption (gray, leftmost). The boundaries of the box are drawn at the first and third quartiles, with the median indicated by a horizontal line. The ``whiskers'' show the extent of the data, with outliers shown as diamond points.}
    \label{fig:boxplot_intTs_v_HCOptype}
\end{figure}

As in \citet{Rybarczyk2022_Obs}, we find that the integrated \hcop{} optical depth --- and therefore, $N(\hcop{})$ and, by extension, $N(\htwo)$ --- decreases with increasing $\langle \ts \rangle$.
In particular, broad (mostly \codarka) \hcop{} absorption is found in regions with systematically higher $\langle \ts \rangle$ (and therefore lower $f_{\mathrm{CNM}}$) than regions with narrow \hcop{} (and narrow CO absorption). This is illustrated more clearly in the box plot shown in Figure \ref{fig:boxplot_intTs_v_HCOptype}. 
These results are consistent with previous work that has shown that denser, CO-bright molecular gas is found preferentially in directions with relatively high CNM fraction \citep[e.g.,][]{SS2014,Nguyen2019,McG2023,Park2023}, and they reinforce the conclusion from \citet{Rybarczyk2022_Obs} that the broad \hcop{} absorption is tracing gas with systematically different physical properties (though \citealt{Rybarczyk2022_Obs} only observed CO emission, not CO absorption).

We estimate $f_{\mathrm{CNM}}$ from $\langle T_s \rangle$ using Equation 18 of \citet{Kim2014} \citep[where we assume a typical CNM cloud temperature between $50~\mathrm{K}$ and $150~\mathrm{K}$ and a typical WNM cloud temperature between $1000~\mathrm{K}$ and $6000~\mathrm{K}$; e.g.,][]{Murray2021}.
Based on the data in Figure \ref{fig:boxplot_intTs_v_HCOptype}, narrow  \hcop{} absorption (associated with narrow CO absorption and, in some cases, CO emission) is detected where $f_{\mathrm{CNM}} = 0.64^{+0.28}_{-0.31}$ while broad (mostly \codarka) \hcop{} absorption is detected where $f_{\mathrm{CNM}} = 0.38^{+0.28}_{-0.27}$ (upper and lower bounds indicate the $15.9$ and $84.1$ percentiles). 

It is also clear from Figure \ref{fig:inttauHCOp_v_intTs} that the \textit{molecular} fraction is systematically lower in regions with broad \hcop{} absorption. For the gas traced by narrow \hcop{} absorption, $f_{\mathrm{mol}}=0.60^{+0.24}_{-0.18}$, while for broad absorption, $f_{\mathrm{mol}}=0.09^{+0.06}_{-0.03}$. In other words, not only is the total amount of \htwo{} lower in these more diffuse regions, but the fraction of hydrogen in \htwo{} is also lower. 
This complements previous work showing that the CO emission detectable in existing surveys arises largely from gas with a high molecular fraction \citep[$\fmol\gtrsim0.5$;][]{Liszt2017}.

\subsubsection{The \hi{} conditions necessary for the formation of molecular gas in the diffuse ISM} \label{subsec:HI_for_molecule_formation}
The association of molecular gas with the CNM is well-demonstrated observationally \citep[e.g.,][]{SS2014}, and well-understood theoretically \citep[e.g.,][]{Goldsmith2007}. In particular, molecular gas has been shown to exist only in directions where the \hi{} optical depth is $\gtrsim0.1$ and the \hi{} spin temperature is $\lesssim150~\mathrm{K}$ \citep{Rybarczyk2022_Obs,Park2023,Hafner2023}.
However, as discussed explicitly by \citet{{Rybarczyk2022_Obs}}, this previous work focused exclusively on narrow absorption --- a fact that is also evident from a qualitative assessment of the published spectra.

Here, despite this difference, our results are largely consistent with previous work, in that we find that most \hcop{} is detected where $\tau_{\rm HI}\gtrsim0.1$, including for the broad \hcop{} absorption. However, while studies focusing on the equivalent of our narrow components find these thresholds to be a necessary but not sufficient condition for the presence of molecules \citep{Hafner2023,Park2023}, our detection rates are far higher --- close to 100\% for $\tau_{\rm HI} > 0.1$. Furthermore, in the case of 3C454.3, we find two \hcop{} components coincident with peak \hi{} optical depths of around $\tau\approx0.01$--$0.05$ (near $v=-16~\kms{}$ and $v=-31~\kms{}$).

\subsubsection{Comment on velocity-resolved properties}

\begin{figure}
    \centering
    \includegraphics[width=\linewidth]{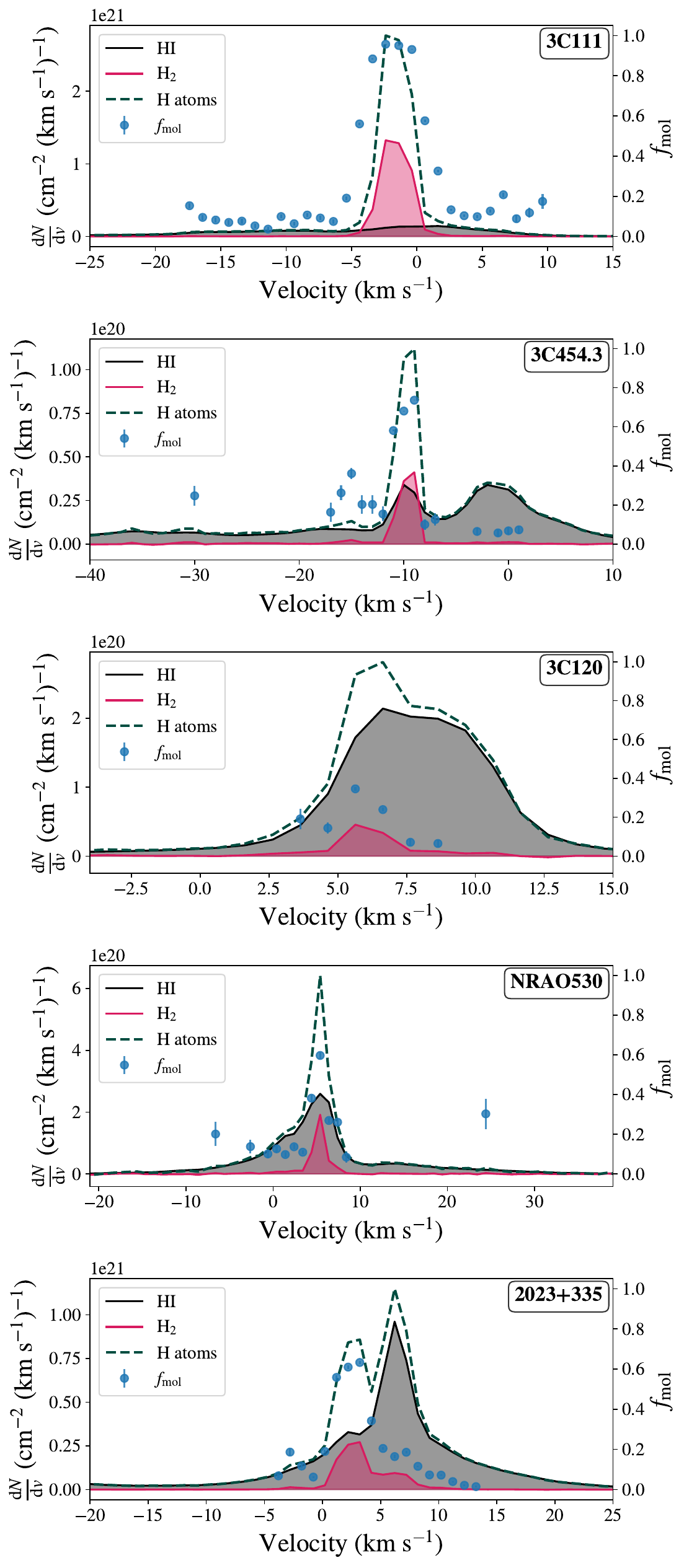}
    \caption{Spectra showing the differential column density in each velocity channel contributed by \htwo{} (pink, calculated from \hcop{}), \hi{} (black, using an isothermal approximation in each channel), and the total hydrogen column (dashed dark green, using the sum of \hi{} and $2\times\htwo{}$). The molecular fraction in each velocity bin where \hcop{} is detected is shown as a blue scatter point. Spectra have been smoothed to $1~\kms{}$ resolution.}
    \label{fig:Av_spectra}
\end{figure}

While we have focused on the integrated properties of \hcop{}, CO, and \hi{}, it is clear from Figures \ref{fig:spectra} and \ref{fig:HI_CO_HCOp_zoomin} that both broad an narrow regions likely contain multiple absorbing components, so the integrated quantities represent weighted averages over multiple clouds.
To look at the atomic and molecular gas more locally, we show the differential column density in each velocity bin in Figure \ref{fig:Av_spectra} (here all spectra have been smoothed to $1~\kms{}$ resolution for clarity).
We further show the estimated molecular fraction in each velocity bin using blue scatter points.
It is clear from Figure \ref{fig:Av_spectra} that the trends reported earlier also hold on a channel-by-channel basis. Typical column densities for the broad components are systematically lower than those for the narrow components, and the same is true of the molecular fraction. 
In future work, we will consider a full spectral decomposition to extend this analysis to a cloud-by-cloud basis (see Section \ref{subsubsec:discussion_comparison_to_OH}).

\section{Discussion} \label{sec:discussion}
Using observations of CO absorption in directions where broad \hcop{} absorption was previously detected, we have shown that the broad \hcop{} absorption signature first identified by \citet{LisztLucas2000} and recently characterized by \citet{Rybarczyk2022_Obs} is indeed mostly \codarkb\ and \codarka\ (Figures \ref{fig:spectra} and \ref{fig:HI_CO_HCOp_zoomin}). The CO absorption observations were necessary to make this comparison, since (1) they probe roughly the same spatial scales as previous \hcop{} absorption observations, and (2) they are more sensitive to low column density CO than emission observations in these directions \citep[our absorption observations have roughly an order of magnitude better column density sensitivity to CO for a given CO abundance compared to the emission data from][]{Dame2022}.  While we detect CO in all directions, nearly all of the CO absorption is associated with strong, narrow \hcop{} absorption.

\subsection{Why is the broad component CO-dark?}
 
\subsubsection{The CO abundance and its implications}

In Section \ref{sec:CO_abundance}, we measure CO abundances (and upper limits) in the diffuse molecular gas probed by our observations (Figure \ref{fig:implied_CO_abundance}), finding values (relative to H$_2$) of $\lesssim10^{-5}$. 
This is low compared to those expected in classic CO-bright molecular clouds $\sim10^{-4}$ \citep[e.g.,][]{Federman1990,Liu2013}. 
However, even for our narrow absorption lines, the gas we are probing is diffuse in almost all cases, as we will discuss presently. The two exceptions are the saturated absorption in 3C111, and the $\sim2.5$ km\,s$^{-1}$ narrow absorption component in J2023+335. The former is associated with the California molecular cloud, shows correspondingly bright CO emission ($\sim3.3~$K) in \citet{Dame2022}, and is in any case excluded from our analysis because of line saturation. The latter is a low Galactic Latitude sightline ($b=-2.36^\circ$) passing through disk molecular gas, with a CO brightness temperature of $\sim1.9~$K \citep{Dame2022}. We will discuss this sightline further below.

For all components, the abundances (as well as the upper limits) that we measure are consistent with the expected values from the literature for diffuse sightlines. For example, the top panel of Figure \ref{fig:implied_CO_abundance} is entirely consistent with Figure 6 of \citet{Sheffer2008}, whose direct observations of CO (optical) and \htwo{} (UV) indicated CO abundances relative to \htwo{} of $\sim10^{-7}$--$10^{-5}$ for $N(\htwo{})\sim10^{19}~\percc{}$--$10^{21}~\percc{}$ --- albeit on a sightline-integrated basis. Meanwhile, it is understood that the CO abundance falls off sharply for $A_V\lesssim1~\mathrm{mag}$ \citep[e.g.,][]{Wolfire2010}. The bottom panel of Figure \ref{fig:implied_CO_abundance} indicates that all of our plotted regions indeed have $A_V\lesssim1~\mathrm{mag}$ \citep[where $A_V \approx N_{\mathrm{H}}/2.1\times10^{21} ~\mathrm{mag}/\persc{}$, e.g.,][]{GuverOzel2009,Zhu2017}, except for the narrow absorption in the direction of J2023+335 (the rightmost point on the plot). 
The CO abundances and upper limits that we measure relative to all hydrogen nuclei are consistent with model expectations for $0.1~\mathrm{mag}\lesssim A_V \lesssim 1~\mathrm{mag}$ from \citet{Wolfire2010}, who modeled CO-dark gas using a 1D photodissocation region with uniform radiation field using code from \citet{Kaufman2006} adapted to better account for CO chemistry. There, while their CO abundances (relative to all hydrogen nuclei) span over five orders of magnitude in this $A_V$ range, they do not exceed $\sim10^{-5}$. 

For the discrepant component --- the J2023+335 narrow absorption --- the implied $A_V > 1~\mathrm{mag}$ and the detection of CO emission might suggest that this component arises from a classic CO-bright molecular cloud, making the low abundance we obtain here puzzling. However, there are several mitigating factors. The first is that the \citet{Dame2022} CO emission data is of very coarse resolution and undersampled, meaning that it may not well represent the pencil beam absorption measurement from which our abundance is derived. The second is that the subthermal excitation temperatures we assume in this work may not be appropriate for this component (if it is indeed associated with emitting gas), meaning that the CO column (and hence the abundance) could be underestimated by a factor of a few. Finally the $A_V$ we estimate here includes a significant contribution from the velocity-integrated \hi{} column within with our broad and narrow velocity ranges. In reality it is clear from the complex, blended \hi{} absorption spectra that multiple components are overlapping in velocity along these sightlines, only some of which may be spatially co-located with molecular gas. As a result, the $A_V$ is almost certainly overestimated to some degree (in all components). 
Decomposing the atomic and molecular absorption spectra in future work (see Section \ref{subsubsec:discussion_comparison_to_OH}) will be important for placing tighter constraints on the CO abundance for individual structures along the line of sight.

\begin{figure}
    \centering
    \includegraphics[width=\linewidth]{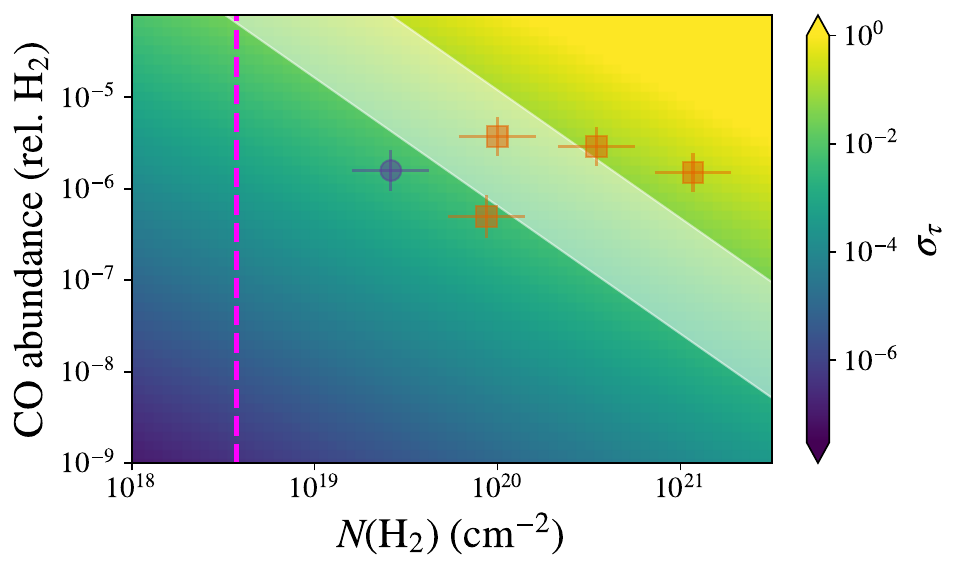}
    \caption{The optical depth sensitivity (color map) required to detect CO at $3\sigma$ for a given \htwo{} column density ($x$-axis) and CO abundance ($y$-axis). The semi-transparent white window shows the theoretical detectability limit for a CO optical depth sensitivity $\sim0.003$ (typical of what we obtain here; see Table \ref{tab:observations}) and a range of full-widths at half-maximum between $1~\kms{}$ and $4~\kms{}$ \citep{LisztLucas1998,Luo2020} and $2.725~\mathrm{K}\leq T_{\mathrm{ex}}\leq6~\mathrm{K}$. The vertical dashed magenta line shows the \htwo{} sensitivity limit for an \hcop{} optical depth sensitivity of 0.002 (see Table \ref{tab:int_tau_CO_HCOp}). CO abundances measured here are shown as semi-transparent points (with the same styles and coloring as in Figure \ref{fig:implied_CO_abundance}, but with non-detections removed). These points are all consistent with the theoretical detectability limit within $<2\sigma$.}
    \label{fig:sensitivity}
\end{figure}

Figure \ref{fig:sensitivity} shows the sensitivity required to detect CO in absorption for a range of \htwo{} column densities and CO abundances appropriate to the diffuse ISM. The approximate sensitivity achieved in the present work is indicated with a semi-transparent window (calculated for $\sigma_\tau\sim3\times10^{-3}$ and assuming a full-width at half-maximum between $1~\kms{}$ and $4~\kms{}$ and excitation temperature $2.725~\mathrm{K}\leq T_{\mathrm{ex}} \leq 6~\mathrm{K}$; \citealt{LisztLucas1998,Luo2020}), consistent with the detections shown in Figure \ref{fig:implied_CO_abundance}.
Based on these model predictions for the CO abundance \citep{Wolfire2010}, and the optical depth sensitivities achievable with current radio observatories, it seems implausible to expect to detect CO in absorption for $N(\htwo{})\lesssim\mathrm{few}\times10^{19}~\mathrm{cm^{-2}}$. This is consistent with our observational results.

While we are focused on CO absorption here, we note that for the \htwo{} column densities ($\lesssim\mathrm{few}\times10^{19}~\persc{}$) and CO abundances ($<10^{-5}$) associated with the broad component (Figure \ref{fig:implied_CO_abundance}), the sensitivity needed to detect CO in emission is $\sigma_{W_{\mathrm{CO}}}\lesssim10^{-2}~\mathrm{K}~\kms{}$ \citep[see, e.g., Equation 18 of][]{Bolatto2013}.
This is below the detection thresholds of even the most sensitive CO emission observations \citep[e.g.,][]{Li2018}. 
Indeed, we detect CO emission (Figure \ref{fig:spectra}) only for the components with $N(\htwo{})>\mathrm{few}\times10^{20}~\persc{}$. This is consistent with theoretical expectations given the column densities and abundances in Figure \ref{fig:implied_CO_abundance} \citep[$W_{\mathrm{CO}}\sim0.2$--$3~\mathrm{K~\kms{}}$; e.g.,][and references therein]{Bolatto2013}.

To summarize, in the diffuse molecular sight lines observed here, the low CO abundances combined with intrinsically low molecular column densities can fully explain why CO is not generally detected with the broad \hcop{} components. Similarly, we need not invoke higher CO abundances in the narrow \hcop{} components to explain their detection in CO --- this can be entirely explained by higher $N(\htwo{})$. CO in emission can be observed in these narrow and moderately diffuse components, despite CO abundances below the canonical 10$^{-4}$.

\subsubsection{The origin of the broad and narrow absorption}
\label{subsubsec:broad_and_narrow_definition}
From Figure \ref{fig:inttauCO_v_inttauHCOp} we see that gas is \codarka\ in our data where $\int\tau_{\hcop{}}dv\lesssim0.2~\mathrm{K~\kms{}}$, corresponding to \htwo{} column densities $\lesssim\mathrm{few}\times10^{19}~\persc$ (as discussed above). The components we identify as narrow are, by definition, those with the highest peak \hcop{} optical depths along the line of sight ($\tau_{\hcop{}}>0.05$), and are therefore more likely to meet this threshold for CO detection. 
This fact remains true even if the CO abundance is the same for gas traced by broad and narrow absorption (a scenario that is not ruled out by the upper limits in Figure \ref{fig:implied_CO_abundance}), because the narrow components are simply where sufficient molecular column has been accumulated to render CO detectable. 
Indeed, in Figure \ref{fig:Av_spectra}, we show that all regions identified as narrow are associated with the highest column densities and molecular fractions in each direction. 

The picture then is one in which the distinction between \textit{broad} and \textit{narrow} absorption in diffuse molecular gas \citep[which we have essentially adopted from previous work, e.g.,][]{LisztLucas2000,Rybarczyk2022_Obs} is primarily one of molecular (H$_2$) column density.  
Along any line of sight, higher column density, better-shielded gas with higher molecular fraction --- and hence sufficient molecular column to be CO-bright (or at least not \codarka) --- will tend to be spatially localized, and hence narrowly confined in velocity. Cold gas at intermediate column densities and lower molecular fraction similarly comprises multiple velocity components, but spread over a wider range of velocities. This gas is also partially molecular, but at a lower level,  accumulating sufficient column density to be detectable in \hcop{} but not in CO. 

In most cases, it seems clear that the breadth of the broad \hcop{} absorption is essentially set by the velocity distribution of the cold gas along the sightline --- the faint \hcop{} absorption traces nearly all velocities where we see \hi{} absorption.
In fact, in some cases (3C111, 3C454.3), our spectra show that the broad \hcop{} absorption clearly comprises multiple discrete components with narrow linewidths, many of which are well-aligned with peaks in the \hi{} absorption spectra (see Section \ref{subsubsec:discussion_comparison_to_OH}).

\subsection{Physical properties of diffuse molecular gas}
\label{subsec:discussion_properties_of_diffuse_molecular_gas}
\subsubsection{Connection to multiphase \hi{}}
\label{subsubsec:discussion_multiphase_HI}
Fundamentally, we are interested in the detection of \codarka\ (and \codarkb) diffuse gas not as an end in itself, but rather as a valuable probe of the physics and chemistry of the diffuse ISM. The multi-wavelength observations presented here have constrained some of the physical conditions of the diffuse CO-dark(*) gas, which is effectively traced by broad \hcop{} absorption.

In Section \ref{sec:results}, we show that narrow absorption features (which are not \codarka, and in some cases are CO-bright) generally align with the optically thickest \hi{} (Figure \ref{fig:spectra}) and the highest total column density (Figure \ref{fig:Av_spectra}) in each direction. Both the CNM fraction and the molecular fraction are relatively high in these environments (Figure \ref{fig:Av_spectra}), despite still being diffuse ($A_V\lesssim1~\mathrm{mag}$) in most cases.

Meanwhile, the broad (mostly \codarka) \hcop{} absorption is associated with lower \hi{} optical depths (Figures \ref{fig:spectra} and \ref{fig:HI_CO_HCOp_zoomin}) and
lower column density gas (Figure \ref{fig:Av_spectra}). 
In at least one case (3C454.3), we detect \hcop{} at velocities where the peak \hi{} optical depth is $\tau\approx0.01$--$0.05$, a factor of 2--10 lower than  previously reported as an empirical threshold for the formation and survival of molecular gas \citep[$\tau\gtrsim0.1$; e.g.,][]{Nguyen2019,Rybarczyk2022_Obs,Park2023,Hafner2023}.
In these environments, the CNM fraction and molecular fraction are both relatively low (Figures \ref{fig:boxplot_intTs_v_HCOptype} and  \ref{fig:Av_spectra}). The broad \hcop{} absorption signature therefore traces gas with systematically different properties to that traced by narrow absorption: this gas is more diffuse, and less of the \hi{} has been converted into either the CNM or into \htwo{}. The conditions of this gas are closer to the threshold of the atomic-to-molecular transition; for example, see Zone C in Figure 3 of \citet{Bellomi2020}, where both the molecular column densities and molecular fraction are comparable to the broad components analyzed in this work. In contrast to \citet{Bellomi2020}, who considered only atomic and molecular column densities integrated along each line of sight, the spectral analysis we present here allows us to isolate the more-diffuse molecular gas from the denser components along a given line of sight and characterize the properties of this transition-state gas directly.

\subsubsection{Comparison to broad, CO-dark OH emission}
\label{subsubsec:discussion_comparison_to_OH}
While we can infer the probable existence of multiple velocity components within the diffuse, \codarka\ (or \codarkb) molecular gas traced by broad \hcop{} absorption, the small-scale spatial structure of this very diffuse molecular gas remains unknown. The observations prevented here cannot distinguish between extended, extremely diffuse gas components of $n(\rm H_2)\sim0.05$\,cm$^{-3}$ \citep[e.g.,][]{LisztLucas2000,Busch2021}, or an ensemble of smaller, denser gas structures \citep[although comparisons to \hcop{} emission do place limits on how dense \hcop{} clumps can be;][]{LucasLiszt1996,LisztLucas2000}.

Given the qualitative similarities between the broad \hcop{} absorption observed by \citet{LisztLucas2000} and \citet{Rybarczyk2022_Obs}, and the broad OH emission observed by \citet{Busch2021}, it is natural to wonder whether these signatures share a common origin. New observations of \hcop{} absorption and OH emission in the same directions will be used to test whether broad \hcop{} absorption and broad OH emission are associated (Busch et al., in prep.). Further, because the OH emission beam is much larger than the \hcop{} absorption pencil-beam, a comparison of the line profiles for the two tracers can be used to constrain the spatial distribution of the diffuse molecular gas producing the broad spectral signatures, thereby providing insight into the structure of the ISM near the atomic-to-molecular transition.

Future work will further use all of these spectral line observations (\hi{} absorption and emission; \hcop{} absorption; and CO absorption and emission; as well as new observations of OH emission and absorption in these directions) to constrain the physical and chemical properties of all diffuse gas clouds detected along each line of sight \citep[see][]{bayes_spec}.
With multiple molecular gas species tracing the broad absorption, we will be able to place tight constraints on the gas kinetic temperature and level of turbulence. From the \hi{} emission and absorption, we will also be able to constrain the temperature of the \hi{} structures associated with broad absorption \citep[previous attempts were unable to constrain the temperature of several components, e.g.,][]{21SPONGE_2018}. 
This decomposition is beyond the scope of this work, and requires upcoming OH emission and absorption observations (Busch et al., in prep.).

\subsection{Association with Galactic structure and kinematics} \label{subsec:kinematics}
As discussed above, one of the defining attributes of \codarkb\ (or \codarka) \hcop{} absorption --- as well as OH emission \citep{Busch2021} --- is the breadth of the profile in velocity \citep{LisztLucas2000,Rybarczyk2022_Obs}. The spectral profile is set by a combination of the local environmental conditions, Galactic rotation, the spatial distribution of gas along the line of sight, and local velocity perturbations. 
While we have discussed the abundance and excitation conditions in detail here, a full description of the origin of the diffuse molecular gas traced by broad \hcop{} absorption will ultimately require constraints on the spatial distribution and kinematics of the diffuse ISM in any given direction. 
For example, does the fact that almost all of the narrow \hcop{} absorption components identified in this work are flanked by broad \hcop{} absorption at both positive and negative velocities (see Figure \ref{fig:HI_CO_HCOp_zoomin}) indicate a physical association of the gas traced by broad and narrow absorption?

Here we attempt to constrain the Galactocentric radii and kinematic distances to absorbing gas structures using the Markov Chain Monte Carlo technique developed by \citet{Wenger2018} implemented in the \texttt{kd} software package\footnote{https://github.com/tvwenger/kd}. We note, though, that the typical uncertainties for both quantities are several hundred parsecs in these directions, (which do not account for the uncertainty in the reliability of kinematic distances at high Galactic latitudes), so it is difficult to conclude anything about the 3D spatial distribution of diffuse molecular gas in detail from our absorption spectra alone.
Moreover, because we cannot map diffuse \hcop{} in absorption and because the filling factor of the diffuse molecular gas traced by broad \hcop{} absorption is unknown, it remains challenging to connect our spectra to structures in the recent 3D maps of the local ISM \citep[e.g.,][]{Edenhofer2024}, except for strongly-absorbing molecular clouds in these directions. For these structures, we find that kinematic distances are consistent within uncertainties, but such uncertainties large (a significant fraction of the distance). 
So while the kinematic distance analysis may provide clues to the large-scale spatial structure of diffuse molecular gas (see below), we argue that more work is needed to constrain the detailed spatial distribution of this gas --- key to understanding its physical origin.

Two of our sightlines --- 3C111 ($\ell=161.8^\circ$) and 3C120 ($\ell=190.4^\circ$) --- point toward the outer Galaxy. At the observed velocities, the gas in the direction of these sources may trace the local arm, the Perseus arm, or perhaps even the Outer arm, with Galactocentric radii $R_{\odot}\lesssim R \lesssim 12~\mathrm{kpc}$. Given that the presence of \hcop{} absorption is nearly continuous across a wide velocity range in each direction, there could also be a contribution from inter-arm gas (in the case of 3C111, at least it is clear that the spectrum comprises far greater than three discrete components).
At a longitude of $\ell=86.1^{\circ}$, the sightline toward 3C454.3 lies nearly parallel to the direction of Galactic rotation. This direction, too, probes mostly gas in the outer Galaxy, with $R_{\odot}\lesssim R \lesssim 10~\mathrm{kpc}$.
These directions all probe gas at slightly lower metallicity \citep[e.g.,][]{Esteban2018,Hawkins2023}.
\citet{Busch2021} showed that significant \codarkb\ gas exists in the outer Galaxy (without absorption, they could not distinguish between \codarkb\ and \codarka), which could be evidence for a widespread thick thick disk ($-200~\mathrm{pc}\lesssim z\lesssim 200~\mathrm{pc}$) of diffuse molecular gas. The near-continuous velocity coverage of \hcop{} absorption we see toward 3C111, 3C120, and 3C454.3 is consistent with widespread diffuse molecular gas at $R_{\odot}\lesssim R \lesssim 12~\mathrm{kpc}$, consistent with the contention of \citet{Busch2021}, but our sample size and lack of spatial information make it difficult to say anything global.

Meanwhile, J2023+335 and NRAO~530 point toward the inner Galaxy. J2023+335 is still reasonably close to the direction of Galactic rotation ($\ell=73.1^\circ$), so despite the kinematic distance ambiguity, the velocities of the observed molecular gas imply that most of this absorption is associated with the local arm and perhaps the Perseus arm. However, NRAO~530 is at low Galactic longitude ($\ell=12.0^\circ$), so this absorption may be local or may come from the far side of the Galaxy \citep[though this is less likely, since at $b=10.1^\circ$, the molecular gas would be at $z\approx300~\mathrm{pc}$ off the plane, well beyond the scale height of dense or diffuse molecular gas, e.g.,][]{Su2021}. In either case, though, the Galactocentric radii are $R_g\approx(7$--$9)~\mathrm{kpc}$.

\section{Conclusions} \label{sec:conclusions}

Because molecular hydrogen is simultaneously the most important molecule in the ISM and extremely difficult to detect in many cold environments, our understanding of the molecular content of galaxies is biased by the imperfect observational approaches we must use to trace \htwo{} indirectly.
Although CO (the second most abundant molecule in galaxies) is often used to trace \htwo{}, the existence of ``CO-dark gas'' --- molecular gas without detectable CO emission (or, in this work, CO absorption) --- is well established, especially in the diffuse ISM. 

Here, we have investigated the atomic and molecular gas properties of \codarkb\ (and the stricter \codarka; Table \ref{tab:categories}) gas associated with a particular spectral signature that has been highlighted in recent literature (\citealt{LisztLucas2000,Rybarczyk2022_Obs,Rybarczyk2023}; see also \citealt{Busch2021}), namely, kinematically broad \hcop{} absorption observed toward diffuse sightlines. 
Our observations of sensitive CO absorption have enabled a direct comparison of the \hcop{} and CO; previous work primarily compared broad \hcop{} absorption to CO emission \citep{Rybarczyk2022_Obs}, which is less sensitive and probes different physical scales than CO pencil-beam absorption observations. We have also used sensitive \hi{} absorption observations \citep[including both new and archival data;][]{21SPONGE_2018} to characterize the multiphase atomic gas in these directions.

We detect almost no CO absorption from the gas traced by broad \hcop{} absorption, while we detect CO absorption associated with all of the narrow \hcop{} features (Figures \ref{fig:spectra} and \ref{fig:HI_CO_HCOp_zoomin}). 
Nevertheless, we do not find evidence that the CO abundance is systematically lower for the gas traced by broad absorption --- the CO abundance (relative to \htwo{}) is $\sim\mathrm{few}\times10^{-7}$ to $\sim10^{-5}$ for narrow absorption, and $\lesssim\mathrm{few}\times10^{-6}$ (upper limits) for the broad absorption. Though significantly lower than the $\sim10^{-4}$ abundance in classic CO-bright molecular clouds \citep[e.g.,][]{Federman1990,Liu2013}, even the narrow absorption traces relatively diffuse gas in almost all cases (see Figures \ref{fig:implied_CO_abundance} and \ref{fig:Av_spectra}), and these abundances are consistent with model predictions \citep{Wolfire2010}.
We further show that, for diffuse molecular gas with with $\text{few}\times10^{18}~\persc{} \lesssim N(\htwo{})\lesssim7\times10^{19}~\persc{}$, \hcop{} absorption is likely to be detected in regions where neither CO emission or CO absorption could be detected with existing facilities (see Figure \ref{fig:sensitivity}).
Thus, in environments close to the atomic-to-molecular transition \citep[e.g.,][]{Bellomi2020}, CO is an ineffective tracer of \htwo{}, while \hcop{} absorption \citep[and perhaps other tracers like OH emission, e.g.,][]{Busch2021, Busch2024} are essential for characterizing this crucial phase in the evolution of gas in galaxies.

The gas traced by broad, \codarka \hcop{} absorption has systematically different properties than the CO-bright (or merely \codarkb) gas in the same directions. The atomic gas coincident with broad \hcop{} absorption has a lower CNM fraction than that associated with narrow \hcop{} absorption (Figure \ref{fig:boxplot_intTs_v_HCOptype}). The fraction of hydrogen in the form of \htwo{} is also lower for the gas associated with broad \hcop{} absorption (Figure \ref{fig:Av_spectra}). As previously suggested by \citet{Rybarczyk2022_Obs}, this could indicate that broad \hcop{} absorption traces gas at the earliest stages of molecule formation in the diffuse ISM. 

Yet, we argue that our observations suggest that the separation of broad and narrow absorption \citep[][]{LisztLucas2000,Rybarczyk2022_Obs} does not reflect a meaningful kinematic difference --- the ``broad'' component simply traces \textit{more diffuse} molecular gas (likely comprised in reality of multiple narrower components) where the \htwo{} column density tends to be lower and CO is thus less likely to be detected in either emission or absorption, but where \hcop{} remains detectable. Indeed, broad and narrow components at similar column densities show similar general properties.

Ultimately, it will be necessary to consider the atomic and molecular gas along each line of sight on a cloud-by-cloud basis.
Future work comparing broad \hcop{} absorption and a similar signature detected in OH emission \citep{LisztLucas2000,Busch2021} may help uncover the nature and physical origin of the broad signature. For example, it remains unclear from this work if the diffuse gas traced by broad \hcop{} emission belongs to a widespread, diffuse phase \citep[e.g.,][]{Busch2021}, or perhaps an ensemble of small, denser clumps. A full spectral decomposition of atomic and molecular observations in these directions will help us to constrain the physical sizes of clouds seen in emission and absorption, as well as characterize their physical and chemical properties on a cloud-by-cloud basis. This work awaits new observations of OH emission (which samples larger physical scales than the \hcop{} pencil-beam) and absorption, which will provide essential alternative probes of the diffuse molecular gas in these directions (Busch et al., in prep.).

\begin{acknowledgements}
The authors acknowledge Interstellar Institute's programs ``II6'' and ``II7'' and the Paris-Saclay University's Institut Pascal for hosting discussions that nourished the development of the ideas behind this work.
D.R.R. is supported by a National Science Foundation Astronomy and Astrophysics Postdoctoral Fellowship under award AST-2303902.
M.P.B. is supported by a Jansky Fellowship provided by the National Radio Astronomy Observatory. The National Radio Astronomy Observatory and Green Bank Observatory are facilities of the U.S. National Science Foundation operated under cooperative agreement by Associated Universities, Inc.
This work is based on observations carried out under project number W21AC and W24BG with the IRAM NOEMA Interferometer. IRAM is supported by INSU/CNRS (France), MPG (Germany) and IGN (Spain).
This paper makes use of the following ALMA data: ADS/JAO.ALMA\#2018.1.00585.S and ADS/JAO.ALMA\#2019.1.01809.S.. ALMA is a partnership of ESO (representing its member states), NSF (USA) and NINS (Japan), together with NRC (Canada), MOST and ASIAA (Taiwan), and KASI (Republic of Korea), in cooperation with the Republic of Chile. The Joint ALMA Observatory is operated by ESO, AUI/NRAO and NAOJ. 
This work benefited from the conference “Structure and polarization in the interstellar medium: A Conference in Honor of Prof. John Dickey”, a hybrid meeting hosted jointly at Stanford University and at the Australia Telescope National Facility in February 2025. We acknowledge support from the National Science Foundation (NSF Award No. 2502957), from the Kavli Institute for Particle Astrophysics and Cosmology, from the Commonwealth Scientific and Industrial Research Organisation, and from the Australian Research Council.
We would also like to thank I. Grenier, M. Gong, and Y. Komichi for useful discussions regarding the content of this paper.
\end{acknowledgements}

\bibliography{refs}{}
\bibliographystyle{aasjournal}

\end{document}